\documentclass[12pt]{article}
\usepackage{bbm}
\usepackage{pifont}
\usepackage{mathrsfs}
\usepackage{amsfonts}
\usepackage{graphicx}
\usepackage{latexsym}
\usepackage{epsfig}
\usepackage{amsmath}
\usepackage{amssymb}
\usepackage{color}
\usepackage{amsthm}
\usepackage[dvipdfm,CJKbookmarks,bookmarksopen=true,colorlinks=true,
linkcolor=blue,citecolor=blue,pdfstartview=FitH,pdftitle=title,pdfauthor=lixx]{hyperref}

\newtheorem{theo}{\sc Theorem}
\newtheorem{lemm}{\sc Lemma}[section]
\newtheorem{rema}{\sc Remark}

\newcommand{\patha}{}

\newcommand{\Cov}{\mbox{Cov}}

\newcommand{\E}{\mbox{E}}

\newcommand{\ve}[1]{\mbox{\boldmath ${#1}$}}
\newcommand{\vesub}[2]{\mbox{{\boldmath ${#1}$}$_{#2}$}}
\newcommand{\vesup}[2]{\mbox{{\boldmath ${#1}$}$^{#2}$}}
\newcommand{\vess}[3]{\mbox{{\boldmath ${#1}$}$_{#2}^{#3}$}}
\newcommand{\hve}[1]{\hat{\ve{#1}}}

\newcommand{\hvesup}[2]{\hat{\ve{#1}}^{#2}}
\newcommand{\hvess}[3]{\hat{\ve{#1}}_{#2}^{#3}}

\newcommand{\bfun}{\left\{\begin{array}{ll}}
\newcommand{\efun}{\end{array}\right.}
\newcommand{\bsmat}{\left[\begin{array}}
\newcommand{\esmat}{\end{array}\right]}
\newcommand{\bmat}{\left(\begin{array}}
\newcommand{\emat}{\end{array}\right)}


\begin{document}
\setlength{\baselineskip}{24pt}
\renewcommand{\baselinestretch}{1.2}
%

\title{Automatic Detection of Significant Areas for Functional Data  with Directional Error Control}
\author{Xu, Peirong \\ Department of Mathematics, Southeast University, China and \\
Department of Statistics, Seoul National University, Korea \\
Lee, Youngjo \\Department of Statistics, Seoul National University, Korea\\
Shi, Jian Qing \thanks{Correspondence to: Dr J. Q. Shi, School of Mathematics \& Statistics, Newcastle University, UK, j.q.shi@ncl.ac.uk.}\\ School of Mathematics \& Statistics, Newcastle University, UK}
\date{}
\maketitle

\noindent {\bf{Abstract:}} To detect differences between the mean curves of two samples in longitudinal study or functional data analysis, we usually need to partition the temporal or spatial domain into several pre-determined sub-areas. In this paper we apply the idea of large-scale multiple testing to find the significant sub-areas automatically in a general functional data analysis framework. A nonparametric Gaussian process regression model is introduced for two-sided multiple tests. We derive an optimal test which controls directional false discovery rates and propose a procedure by approximating it on a continuum. The proposed procedure controls directional false discovery rates at any specified level asymptotically. In addition, it is computationally inexpensive and able to accommodate different time points for observations across the samples. Simulation studies are presented to demonstrate its finite sample performance. We also apply it to an executive function research in children with Hemiplegic Cerebral Palsy and extend it to the equivalence tests.\vspace{9pt}

\par

\fontsize{10.95}{14pt plus.8pt minus .6pt}\selectfont

\section{Introduction}

\begin{figure}[htbp]
\centering \resizebox{12cm}{8cm}{\includegraphics{\patha 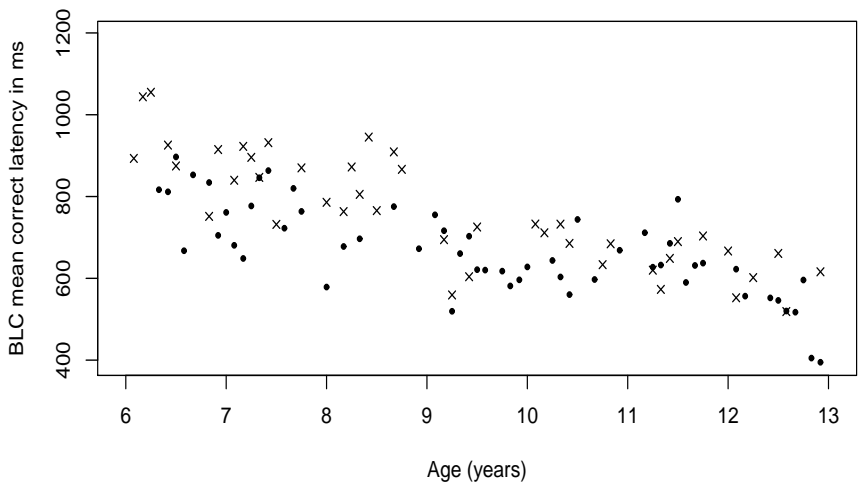}}
\caption{The scatterplot of BLC mean correct latency data in AVGPs group ($\centerdot$) and NAVGPs group ($\times$). }\label{fig1}
\end{figure}

The testing problem in functional data analysis framework  is motivated by an example on studying executive functions  in children with Hemiplegic Cerebral Palsy. The Big/Little Circle (BLC) test is an attention measure that tests comprehension, learning and reversal of a rule (see e.g. Moore and Puri, 2012). In this study, the data on BLC mean correct latency was collected from 141 students, aging from 6 to 13, who completed the BLC test. Among them, 56$\%$ are action video game players (AVGPs) and 44$\%$ are non-action video game players (NAVGPs) as shown in Figure~\ref{fig1}. Let $Y_1(t)$ and $Y_2(t)$ be the BLC mean correct latency for NAVGPs  and AVGPs groups respectively, where $t$ is the age of children. They are continuous functional variables although observations are collected at discrete points.   We are interested in identifying ages that the means of $Y_1(t)$ and $Y_2(t)$  have significant difference. In particular, we wish to detect the specific areas of age where the significant differences occur. We will refer such areas as {\it significant areas}. Thus we wish to detect the significant areas automatically and at the same time minimize the false nondiscovery rate while controlling false discovery rates.

Functional data analysis (FDA) has emerged as a popular area of statistics over the last decade for the analysis of data with functional features, such as growth curves, motion and image data. Ramsay and Silverman (2005) and Ramsay et al. (2009) offered applied-oriented introductions to the ideas and tools of FDA. Ferraty and Romain (2011) reviewed some recent theoretical developments of FDA. Other important directions related to statistical inference in FDA includes Bosq (2000), Yao et al. (2005), M\"uller (2005), Ferraty and Vieu (2006), Di et al. (2009), Horv$\acute{a}$th and Kokoszka (2012), and Wang and Shi (2014) among many others. However, hypothesis testing with directional error control on detecting areas in which differences of the mean curves of two samples are significant (i.e. detecting the {\it significant areas}) has received little attention. Inspired by the recent development of large-scale multiple testing for complex big data (see e.g. Zhang et al., 2011, Lee and Lee, 2014 and Sun et al. 2015), we propose an automatic detection procedure to find significant areas and allow control of the directional error at the same time.

Testing differences in the mean functions of two samples of curves has been approached in many literatures. For example, Zhang et al. (2010) introduced an $L^2$-norm based test, Horv$\acute{a}$th et al. (2013) developed a test based on the sample means of the curves, and Staicu et al. (2014) proposed a pseudo likelihood ratio test. Extension to multiple samples of curves was discussed in Shen and Faraway (2004). Cuevas et al. (2004), Est$\acute{e}$vez-P$\acute{e}$rez and Vilar (2008) and Cuesta-Albertos and Febrero-Bande (2010) further extended it to the functional analysis of variance. Those works all focused on  detecting the overall difference. However, we are often interested in determining the sub-areas of the functional domain (temporal or spatial) where  the mean curves are significant different in many problems such as the motivating example we discussed earlier. To identify specific areas for a significant difference, Ramsay and Silverman (2005) proposed a pointwise t-test without multiplicity control, and Cox and Lee (2008) applied the Westfall-Young randomization method to control the family-wise error rate (FWER). However, when the number of null hypotheses is large, lack of multiplicity control is too permissive, while the full protection resulting from controlling the FWER is too stringent.

Compared with FWER in the context of multiple testing, the false discovery rate (FDR) introduced by Benjamini and Hochberg (1995) has received great attention during the past decade. Lots of procedures have been proposed in large-scale scientific studies with goals of controlling the FDR. For instance, Benjamini and Hochberg (1995) provided a sequential p-value method to control FDR; Sun and Cai (2009) introduced an asymptotical optimal procedure with test statistics under dependence; Liu et al. (2012) proposed a graphical-model based multiple testing procedure to genome-wide association studies; Lee and Bj$\o$rnstad (2013) expressed the problem of multiple testing as an inference problem with basic responses. Other relevant works are Storey (2002), Efron (2004, 2007), Genovese and Wasserman (2004), Zhang et al. (2011), French and Sain (2013) and some of the references therein. When the tests are two-sided as in our motivating example, it often becomes essential for researchers to further determine the direction of significance, rather than significance alone. Then, the decisions can potentially lead to three types of errors for each test: Type I error if the null hypothesis is true but rejected, Type II error if the null hypothesis is not true but failed to reject, and Type III error if the null hypothesis is not true but the direction of the alternative is falsely declared. To deal with Type I as well as Type III errors in the FDR framework, Benjamini and Yekutieli (2005) proposed a  so-called directional Benjamini-Hochberg (BH) procedure for independent tests, Guo et al. (2010) extended the directional BH procedure based on the Bonferroni test to gene expression data with ordered categories, Clements et al. (2014) introduced a three-stage directional BH procedure to study vegetation fluctuations, and Lee and Lee (2014) developed an optimal extended likelihood test with directional FDRs under hidden Markov random field models. However, the multiple testing problems mentioned above are all restricted to the assumption that each hypothesis has its own observed data, while in our motivating example, we only observed BLC mean current latency at finite time points in age range of [6, 13] but we need to make decisions at any age (time) between 6 and 13. Recently, Sun et al. (2015) developed an asymptotic optimal data-driven procedure that controls the FDR for multiple testing on a continuous domain, where the optimality is restricted in a set that test statistic satisfies monotone ratio condition. Their method is confined to change detection of one curve that may not be applicable to test differences in the two mean curves. And they derived the oracle procedure for two-sided tests by only controlling the FDR related to Type I error, which implies that their method may not be powerful in multiple tests with more than two actions.

To address the issue, we propose a new directional FDR procedure for detecting differences in the mean functions of two samples of functional data observed at discrete grid points. This would be the first attempt to handle two-sample multiple testing for detecting mean differences by controlling FDR in functional data analysis framework. In contrast to pointwise testing idea, we introduce a nonparametric Gaussian process regression model for directional two-sided multiple tests. It provides a natural framework on modeling mean structure and covariance structure of the difference between two curves simultaneously and the latter can be used to effectively extract information from nearby points for decision making. In the spirit of definitions in discrete cases, we define the directional FDRs for the continuous hypothesis testing process, and derive a test which optimally controls directional FDRs among all decision rules for multiple testing. Further, to make the continuous decision process applicable, a procedure is proposed by approximating the optimal test on a continuum. It is shown that it can control directional FDRs at any specified level asymptotically. 
Compared with conventional methods, our simulation studies manifest the drastically improved performance of the proposed procedure on directional error control and power.

The rest of the paper is organized as follows. In Section~\ref{sec2}, we formulate the multiple testing problem and introduce directional FDRs for this continuous hypothesis testing process. Section~\ref{sec3} derives the optimal test with directional FDRs and presents a procedure for implementation. In Section~\ref{sec4}, we investigate the finite sample performance of the proposed procedure by simulation studies and an application to the executive function study. The method is extended to equivalence tests in Section~\ref{sec7}. The paper is concluded with a discussion in Section~\ref{sec5} and all the technical details are relegated to Appendix.

\section{Problem formulation and directional FDRs}\label{sec2}

In this section, we formulate the multiple testing problem of detecting differences in the mean functions of two samples of curves and introduce directional FDRs on a continuum.

Let $Y_g(t), g=1,2$ be two curves of functional data, which are functions of $t$. In functional data analysis, $t$ denotes a real-valued variable, which could be time or some other temporal or spatial variable. In this paper, without loss of generality, we assume $t$ is time as in our motivating example and the corresponding time range is a closed interval $T$; for simplicity take $T=[0,1]$. We are interested in detecting differences between $\E(Y_1(t))$ and $\E(Y_2(t))$ over time on $T$. Specifically, consider the following functional regression model
\begin{eqnarray}\label{model}
Y_1(t) &=& \mu(t) + \mu_d(t) + \epsilon_1(t),\nonumber\\
Y_2(t) &=& \mu(t) + \epsilon_2(t),
\end{eqnarray}
where $\mu(\cdot)$ and $\mu_d(\cdot)$ are unknown functions and $\epsilon_1(t)$ and $\epsilon_2(t)$ are the independent random errors. Then, for each time $t$, we are interested in the directional two-sided test
\begin{eqnarray}\label{test}
& & H_0(t):   |\mu_d(t)| \leq \Delta \nonumber \\
&\mbox{versus}&  H_1(t): \mu_d(t) < -\Delta\quad \mbox{or} \quad H_2(t): \mu_d(t) > \Delta,
\end{eqnarray}
where $\Delta$ is a pre-specified constant, denoting the size of difference we are interested in. Assume that there is an underlying state $z(t)$ associated with each time $t$ taking one of three states. We set $z(t)=0$ if hypothesis at time $t$ is the null and $z(t)=1$ or 2 if hypothesis at time $t$ is the alternative 1 or 2, respectively. Let $\delta(t)\in \{0,1,2\}$ be a decision rule for the hypothesis $H_0(t)$. If $\delta(t)=z(t)$, the hypothesis is correctly identified by the decision rule, otherwise there exist errors. Let $R_k=\{t\in T: \delta(t)=k\}$ and $V_{jk}=\{t \in T: z(t)=j, \delta(t)=k\}$ for $j,k=0,1,2$. Table~\ref{tab1} summarizes the possible outcomes of multiple testing with two alternatives, which shows that there exist three types of errors in the directional two-sided multiple testing \eqref{test}.

\begin{center}{\small
\begin{table}[htb!]
\caption{\label{tab1} {Outcomes of multiple testing with two alternatives} \hspace{6.5cm}}{\small {\footnotesize \tabcolsep
0.09cm
\renewcommand{\arraystretch}{1.2}
\begin{tabular}{l|ccc|c}
\hline
& Declared as null & Declared as alternative 1 & Declared as alternative 2 & Total\\
& $\delta(t)=0$ & $\delta(t)=1$ & $\delta(t)=2$ & \\ \hline
Null ($z(t)=0$)& $V_{00}$     &  $V_{01}$ (\emph{Type I error}) & $V_{02}$ (\emph{Type I error}) & $T_0$\\
Alternative 1 ($z(t)=1$) &   $V_{10}$ (\emph{Type II error}) & $V_{11}$ & $V_{12}$ (\emph{Type III error}) & $T_1$\\
Alternative 2 ($z(t)=2$) &   $V_{20}$ (\emph{Type II error}) & $V_{21}$ (\emph{Type III error}) & $V_{22}$ & $T_2$\\
\hline
Total & $R_0$ & $R_1$  & $R_2$ & $T$\\
\hline
\end{tabular}%
}}
\end{table}
}
\end{center}

Let $\mathcal{L}(\cdot)$ be the Lebesgue measure on time range $T$. Then, $\mathcal{L}(N_1)=\mathcal{L}(V_{01})+\mathcal{L}(V_{02})$ and $\mathcal{L}(N_2)=\mathcal{L}(V_{10})+\mathcal{L}(V_{20})$ are the sizes of areas corresponding to Type I and Type II errors, respectively, and $\mathcal{L}(N_3)=\mathcal{L}(V_{12})+\mathcal{L}(V_{21})$ is the size of area corresponding to Type III error, a directional error. When the interest is to test hypotheses at individual time points, a natural and practical way is to control an error rate in the FDR framework by considering all of these three types of errors. Thus, in this paper, we propose to control either the sum of Type I and Type III errors while minimizing the Type II error or control the Type I error while minimizing the sum of Type II and Type III errors. Let $a \vee b = \max\{a, b\}$. Define FDR and the marginal FDR (mFDR) for Type I error rate as
\begin{equation*}
\textup{FDR}_{\textup{I}} = \E\left\{\frac{\mathcal{L}(N_1)}{\mathcal{L}(R_1\cup R_2)\vee 1}\right\}\ \mbox{and}\ \textup{mFDR}_{\textup{I}} = \frac{\E\{\mathcal{L}(N_1)\}}{\E\{\mathcal{L}(R_1\cup R_2)\}},
\end{equation*}
those for Type III error rate as
\begin{equation*}
\textup{FDR}_{\textup{III}} = \E\left\{\frac{\mathcal{L}(N_3)}{\mathcal{L}(R_1\cup R_2)\vee 1}\right\}\ \mbox{and}\ \textup{mFDR}_{\textup{III}} = \frac{\E\{\mathcal{L}(N_3)\}}{\E\{\mathcal{L}(R_1\cup R_2)\}},
\end{equation*}
and those for the sum of the Type I and Type III error rates as
\begin{equation*}
\textup{FDR}_{\textup{I+III}} = \E\left\{\frac{\mathcal{L}(N_1\cup N_3)}{\mathcal{L}(R_1\cup R_2)\vee 1}\right\}\ \mbox{and}\ \textup{mFDR}_{\textup{I+III}} = \frac{\E\{\mathcal{L}(N_1\cup N_3)\} }{\E\{\mathcal{L}(R_1\cup R_2)\}}.
\end{equation*}
Besides the error rate for discoveries, we can define similar error rate for false nondiscoveries, the false nondiscovery rate and the marginal false nondiscovery rate
\begin{equation*}
\textup{FNDR}=\E\left\{\frac{\mathcal{L}(N_2)}{\mathcal{L}(R_0)\vee 1}\right\}\ \mbox{and}\ \textup{mFNDR} = \frac{\E\{\mathcal{L}(N_2)\}}{\E\{\mathcal{L}(R_0)\}},
\end{equation*}
which is related to Type II error. Further, to compute the power of a single directional two-sided testing procedure, Leventhal and Huynh (1996) recommended excluding Type III error from the conventional power. Therefore, in this paper, we define a modified power (MP) of a directional two-sided multiple testing procedure by considering both Type II and Type III errors
\begin{equation*}
\textup{MP} = 1 - \frac{\E\{\mathcal{L}(N_2 \cup N_3)\}}{\E\{\mathcal{L}(T_1 \cup T_2)\}}.
\end{equation*}

{\rema Lee and Lee (2014) created a similar table to summarize the outcomes of multiple testing with two alternatives and defined the corresponding directional FDRs. The key difference here is that for continuous testing process \eqref{test}, the false discovery measures are related to the sizes of areas corresponding to three types of errors, which couldn't be calculated directly by counting the number of cases as in discrete case where each hypothesis has its own observed data. Therefore, a new strategy is needed to develop for inference based on the continuous functional data analysis framework but using the data observed at discrete points.}

\section{Optimal tests for automatic detection of significant areas}\label{sec3}

\subsection{Optimal procedures for controlling directional FDRs}\label{sec3-1}

Suppose the observed data $\{(Y_{1i},t_{1i}): i=1,\ldots,n_1\}$ and $\{(Y_{2i},t_{2i}): i=1,\ldots,n_2\}$ are realizations of two underlying stochastic processes from model \eqref{model}. The notation of the time points, $t_{1i}$ and $t_{2i}$, allows for different observation points in the two groups, and $t_{1i}$'s and $t_{2i}$'s consist of subsets of $T$. Our objective is to predict the states of hypothesis $z(t) \in \{0,1,2\}$ at any time point $t \in T$ in an optimal way. Therefore, it is necessary to exploit the temporal correlations and extract information from nearby points for prediction. Consider a loss function
\begin{equation}\label{loss}
L(\delta,z;\lambda) = \lambda_1\mathcal{L}(N_1) +\lambda_2\mathcal{L}(N_2) + \lambda_3\mathcal{L}(N_3),
\end{equation}
where $\lambda_1,\lambda_2$ and $\lambda_3$ are relative costs. The following theorem derives the optimal rule for the weighted classification problem \eqref{loss}.

{\theo\label{theo1} Let $\mathcal{D}$ be the whole data set consisting of $\{(Y_{1i},t_{1i}): i=1,\ldots,n_1\}$ and $\{(Y_{2i},t_{2i}): i=1,\ldots,n_2\}$. Assume all parameters in model \eqref{model} are known. Then,
\begin{itemize}
\item[(1)] if $\lambda_2=1$ and $\lambda_1=\lambda_3=\lambda$ in \eqref{loss}, the optimal decision rule $\delta^{(I+III)}=\{\delta^{(I+III)}(t): t\in T\} = \textup{argmin}_{\delta} \E\{L(\delta,z;\lambda)|\mathcal{D}\}$ becomes
\begin{eqnarray*}
\delta^{(I+III)}(t) &=& 2\ \mbox{if}\ \frac{1-{\rm P}(z(t)=0\mid \mathcal{D})}{1-{\rm P}(z(t)=2\mid \mathcal{D})} > \lambda\ \mbox{and}\ {\rm P}(z(t)=2\mid \mathcal{D}) > {\rm P}(z(t)=1\mid \mathcal{D}),\\
&=& 1\ \mbox{if}\ \frac{1-{\rm P}(z(t)=0\mid \mathcal{D})}{1-{\rm P}(z(t)=1\mid \mathcal{D})} > \lambda\ \mbox{and}\ {\rm P}(z(t)=2\mid \mathcal{D}) \leq {\rm P}(z(t)=1\mid \mathcal{D}),\\
&=& 0\ \mbox{otherwise};
\end{eqnarray*}
\item[(2)] if $\lambda_1=\lambda$ and $\lambda_2=\lambda_3=1$ in \eqref{loss}, the optimal decision rule $\delta^{(I)}=\{\delta^{(I)}(t): t\in T\} = \textup{argmin}_{\delta} \E\{L(\delta,z;\lambda)|\mathcal{D}\}$ becomes
\begin{eqnarray*}
\delta^{(I)}(t) &=& 2\ \mbox{if}\ \frac{{\rm P}(z(t)=2\mid \mathcal{D})}{{\rm P}(z(t)=0\mid \mathcal{D})} > \lambda\ \mbox{and}\ {\rm P}(z(t)=2\mid \mathcal{D}) > {\rm P}(z(t)=1\mid \mathcal{D}),\\
&=& 1\ \mbox{if}\ \frac{{\rm P}(z(t)=1\mid \mathcal{D})}{{\rm P}(z(t)=0\mid \mathcal{D})} > \lambda\ \mbox{and}\ {\rm P}(z(t)=2\mid \mathcal{D}) \leq {\rm P}(z(t)=1\mid \mathcal{D}),\\
&=& 0\ \mbox{otherwise}.
\end{eqnarray*}
\end{itemize}}

Theorem~\ref{theo1} gives the optimal rules for various weighted classification problems. We next show that the optimality property can be extended to the multiple testing problems with respect to various directional FDRs defined in Section~\ref{sec2}.

{\theo\label{theo3} Let $\mathcal{A}=\{\delta^{(I+III)}: \lambda>0\}$ be the collection of decision rules in form of $\delta^{(I+III)}$ derived in Theorem~\ref{theo1}. Given an mFDR$_{\textup{I+III}}$ level $\alpha$, let $\delta=\{\delta(t): t \in T\}$ be any decision rule satisfying mFDR$_{\textup{I+III}}\{\delta\}\leq \alpha$. Then, there exists a $\lambda$ determined by $\delta$ such that $\delta^{(I+III)} \in \mathcal{A}$ performs better than $\delta$ in the sense that
\begin{equation*}
\textup{mFDR}_{\textup{I+III}}\{\delta^{(I+III)}\} \leq \textup{mFDR}_{\textup{I+III}}\{\delta\} \leq \alpha,
\end{equation*}
and
\begin{equation*}
\textup{mFNDR}\{\delta^{(I+III)}\} \leq \textup{mFNDR}\{\delta\}.
\end{equation*}}

Theorem~\ref{theo3} demonstrates that the optimal decision rule for controlling the sum of Type I and Type III errors with the smallest Type II error belongs to the set $\mathcal{A}$. In other words, one only needs to search in $\mathcal{A}$ for the optimal rule, instead of searching for all decision rules. Similarly, it can be shown that the optimal decision rule for controlling Type I error with the smallest sum of Type II and Type III errors is in the form of $\delta^{(I)}$ derived in Theorem~\ref{theo1}.

{\rema \label{remaTh2} In the case of $\lambda_3=0$, Sun et al. (2015) showed the optimal solution to the weighted classification problem is optimal in $\{\delta: \delta(t)=I\{T(t)<c\}, T\ \mbox{statisfies monotone ratio condition}\}$ for the multiple testing problem, but Theorem~\ref{theo3} extends the result to a more general case, revealing that this solution is even optimal among all decision rules for the multiple testing.}

\subsection{Extension to practical situations}\label{sec3-2}

It is not straightforward to use the optimal procedures described in Section~\ref{sec3-1} because (a) it is impossible to make an uncountable number of decisions on $T$, and (b) the true smooth trajectories $\mu(\cdot)$ and $\mu_d(\cdot)$ are not directly observable and thus the test statistics should be evaluated at unobserved time points. In this section, we develop procedures for directional FDRs control to overcome these difficulties.

To address (a), we first divide the interval $T=[0,1]$ into $N$ equal-length subintervals $[s_{i-1}, s_i)$ with $s_0=0$ and $s_i=s_{i-1}+1/N$, $i=1,\ldots,N-1$, and pick the center point $t^*_i$ in $[s_{i-1}, s_i)$, $i=1,\ldots,N-1$. Then, for a decision rule $\delta$, we have
\begin{eqnarray*}
E\{\mathcal{L}(N_1)\} &=& \int^1_0 \E\{I(\delta(t)\neq 0){\rm P}(z(t)=0|\mathcal{D})\}d\mathcal{L}(t) =
\lim_{N\rightarrow \infty} \frac{1}{N}\sum^N_{i=1}\E\{S_0(t^*_i)I(\delta(t^*_i)\neq 0)\},\\
E\{\mathcal{L}(N_3)\} &=& \lim_{N\rightarrow \infty} \frac{1}{N}\sum^N_{i=1}\E\{S_2(t^*_i)I(\delta(t^*_i)=1)+S_1(t^*_i)I(\delta(t^*_i)=2)\},\ \mbox{and}\\
E\{\mathcal{L}(N_1\cup N_3)\} &=& \lim_{N\rightarrow \infty} \frac{1}{N}\sum^N_{i=1}\sum^2_{k=1}\sum_{j\neq k}\E\{S_j(t^*_i)I(\delta(t^*_i)=k)\}\\
&=& \lim_{N\rightarrow \infty} \frac{1}{N}\sum^N_{i=1}\sum^2_{k=1}\E\{I(\delta(t^*_i)=k)(1-S_k(t^*_i))\},
\end{eqnarray*}
where function $S_k(t)= {\rm P}(z(t)=k|\mathcal{D})$, $k=0,1,2$. Therefore, motivated by the limit definition of a definite integral, mFDR$_{\textup{I}}$ can be estimated by
\begin{equation}\label{mFDR1}
\widehat{\textup{mFDR}}_{\textup{I}}(\lambda)=\frac{1}{r}\sum^N_{i=1}S_0(t^*_i)I(\delta(t^*_i)\neq 0)
\end{equation}
for any given $\lambda$ and all parameters in model \eqref{model}, where $r=\sum^N_{i=1}I(\delta(t^*_i)\neq 0)$. According to Theorem~\ref{theo1}, it is easy to see that $\delta^{(I)}(t)=0$ if $S_0(t) \leq (1+\lambda)^{-1}$. Suppose that $\lambda_1$ and $\lambda_2$ are chosen so that $h_{(r)} < (1+\lambda_2)^{-1} < h_{(r+1)} < (1+\lambda_1)^{-1} < h_{(r+2)}$, where $h_{(r)}$ is the $r$th smallest value of $S_0(t^*_i)$. Then,
\begin{eqnarray*}
\widehat{\textup{mFDR}}_{\textup{I}}(\lambda_2) - \widehat{\textup{mFDR}}_{\textup{I}}(\lambda_1) &=& r^{-1}\sum^r_{i=1}h_{(i)} - (r+1)^{-1}\sum^{r+1}_{i=1}h_{(i)} \\
&=& \{r(r+1)\}^{-1}\{\sum^r_{i=1}h_{(i)} - rh_{(r+1)}\} < 0.
\end{eqnarray*}
Thus, $\widehat{\textup{mFDR}}_{\textup{I}}$ monotonically decreases with $\lambda$, and we propose the following step-down test procedure for FDR$_{\textup{I}}$ control:
\begin{eqnarray}\label{mFDRtest}
\mbox{let}\ \lambda^* &=& \inf\{\lambda: \widehat{\textup{mFDR}}_{\textup{I}}(\lambda) \leq \alpha\};\ \mbox{then}\nonumber\\
\delta^{(I)}(t) &=& \sum^N_{i=1}I(s_{i-1} \leq t < s_i)\delta^{(I)}(t^*_i)\\
\mbox{with}\ \delta^{(I)}(t^*_i)&=& 2\ \mbox{if}\
\frac{S_2(t^*_i)}{S_0(t^*_i)} > \lambda^*\ \mbox{and}\ S_2(t^*_i) > S_1(t^*_i),\nonumber\\
&=& 1\ \mbox{if}\ \frac{S_1(t^*_i)}{S_0(t^*_i)} > \lambda^*\ \mbox{and}\ S_2(t^*_i) \leq S_1(t^*_i),\nonumber\\
&=& 0\ \mbox{otherwise}.\nonumber
\end{eqnarray}
The following theorem shows that this test controls FDR$_{\textup{I}}$ at level $\alpha$ asymptotically, which implies that the proposed procedure \eqref{mFDRtest} approximates a multiple comparison correction for a continuous comparison process \eqref{test} as the grid for pointwise comparisons becomes finer.

{\theo\label{theo2} Let $\{\cup^N_{i=1}[s_{i-1},s_i): N=1,2,\ldots\}$ be a sequence of partitions of $T$ satisfying Conditions C1 and C2 in the Appendix. Then, the $\textup{FDR}_{\textup{I}}$ level of procedure \eqref{mFDRtest} satisfies $\textup{FDR}_{\textup{I}} \leq \alpha + o(1)$ when $N \rightarrow \infty$. }

{\rema \label{rema3} For simplicity, we choose the center point $t^*_i$ in each subinterval $[s_{i-1},s_i)$ as a representative point. But from the proof of Theorem~\ref{theo2}, we can see that, no matter which point is chosen as a representative point in $[s_{i-1},s_i)$, the proposed procedure \eqref{mFDRtest} controls FDR$_{\textup{I}}$ at the nominal level asymptotically as long as Conditions C1 and C2 are fulfilled. }

Similarly, by using $\widehat{\textup{mFDR}}_{\textup{I+III}}(\lambda)= \frac{1}{r}\sum^N_{i=1}\sum^2_{k=1}I(\delta(t^*_i)=k) (1-S_k(t^*_i))$, we control FDR$_{\textup{I+III}}$ at the nominal level. However, they are still difficult to implement because of (b).

Further to address (b), we propose a Gaussian process regression (GPR) model for \eqref{model} to estimate unknown quantities $S_k(t)= {\rm P}(z(t)=k|\mathcal{D})$, $k=0,1,2$. GPR model is a good choice as a globally approximated nonlinear functional regression model in \eqref{model} (in contrast with locally approximated model for most of conventional nonparametric model); see the details in Shi and Choi (2011) and Wang and Shi (2014). Specifically, consider $\{\mu(t): t \in T\}$ and $\{\mu_d(t): t \in T\}$ as independent random processes and suppose they have Gaussian process priors with zero means and kernel functions $\kappa(\cdot,\cdot;\ve{\eta})$ and $\gamma(\cdot, \cdot; \ve{\theta})$, respectively, where $\Cov(\mu(t),\mu(t'))=\kappa(t,t';\ve{\eta})$ and $\Cov(\mu_d(t), \mu_d(t'))=\gamma(t,t';\ve{\theta})$. Assume that $\{\epsilon_1(t): t\in T\}$ and $\{\epsilon_2(t): t\in T\}$ are Gaussian white noise processes with zero mean and variance $\sigma^2$, which are independent from each other and to both $\{\mu(t): t \in T\}$ and $\{\mu_d(t): t \in T\}$. One example of the kernel function $\gamma(\cdot,\cdot;\ve{\theta})$ is the following squared exponential covariance function with a nonstationary linear term:
\begin{equation}\label{gamma}
\gamma(t_i,t_j;\ve{\theta})=\xi\exp\left\{-\omega(t_i-t_j)^2/2\right\}+\zeta t_i t_j,
\end{equation}
where $\ve{\theta}=(\xi,\omega,\zeta)$ is a set of hyper-parameters. When $\zeta=0$, the kernel function $\gamma(\cdot,\cdot;\ve{\theta})$ reduces to so-called squared exponential covariance function, which is stationary and nondegenerate (Rasmussen and Williams, 2006). The parameter $\omega$ corresponds to the smoothing parameters in spline. So, we call $\omega^{-1}$ the length-scale. A large length-scale implies the underlying curve is expected to be essentially flat and the decrease in length-scale results in more rapidly fluctuating functions.

Let $n=n_1+n_2$, $\ve{\Theta}=(\vesup{\eta}{T}, \vesup{\theta}{T}, \sigma^2)^T$, and  $\ve{Y}=(\vess{Y}{1}{T}, \vess{Y}{2}{T})^T$ with $\vesub{Y}{1}=(Y_{11},\ldots,Y_{1n_1})^T$ and $\vesub{Y}{2}=(Y_{21},\ldots,Y_{2n_2})^T$. Consider the joint density function of $\ve{Y}, \ve{\tilde{\mu}}$ and $\vesub{\tilde{\mu}}{d}$
\begin{equation}
f_{\ve{\Theta}}(\ve{Y},\ve{\tilde{\mu}},\vesub{\tilde{\mu}}{d}) = \phi(\ve{\tilde{\mu}}\mid \ve{0},\vesub{K}{n}) \phi(\vesub{\tilde{\mu}}{d}\mid \ve{0},\vesub{\Gamma}{n_1})\prod^{n_1}_{i=1}\phi(Y_{1i}\mid \mu(t_{1i})+\mu_d(t_{1i}), \sigma^2) \prod^{n_2}_{i=1}\phi(Y_{2i}\mid \mu(t_{2i}), \sigma^2),
\label{hlik}
\end{equation}
where $\ve{\tilde{\mu}}=(\mu(t_{11}),\ldots,\mu(t_{1n_1}), \mu(t_{21}),\ldots,\mu(t_{2n_2}))^T$, $\vesub{\tilde{\mu}}{d}=(\mu_d(t_{21}),\ldots,\mu_d(t_{2n_2}))^T$, $\phi(\cdot)$ is the density of (multivariate) normal distribution, $\vesub{K}{n}$ and $\vesub{\Gamma}{n_1}$ are covariance matrices of $\ve{\tilde{\mu}}$ and $\vesub{\tilde{\mu}}{d}$, respectively, with $(i,j)$th element $\kappa(t_i,t_j;\ve{\eta})$ and $\gamma(t_i,t_j;\ve{\theta})$. Then, the parameters $\ve{\Theta}$ can be consistently estimated by maximizing the likelihood (see Shi and Choi, 2011)
\begin{equation*}
l(\ve{\Theta};\ve{Y})=f_{\ve{\Theta}}(\ve{Y})=\int \int f_{\ve{\Theta}}(\ve{Y},\ve{\tilde{\mu}},\vesub{\tilde{\mu}}{d}) d\ve{\tilde{\mu}}d\vesub{\tilde{\mu}}{d}.
\end{equation*}
Let $\hve{\Theta}=(\hvesup{\eta}{T}, \hvesup{\theta}{T}, \hat{\sigma}^2)^T$ be the estimates of $\ve{\Theta}$. Then, we can make inference about $\mu_d(t)$ by using $f_{\hve{\Theta}}(\mu_d(t)\mid \mathcal{D})$. As the sample size $n$ goes to infinity, we have $f_{\hve{\Theta}}(\mu_d(t)\mid \mathcal{D}) \rightarrow f_{\ve{\Theta}}(\mu_d(t)\mid \mathcal{D})$.

Now, we consider how to make inference about $\vess{\mu}{d}{N}=(\mu_d(t^*_1),\ldots,\mu_d(t^*_N))^T$, where $\vesup{T}{*}=(t^*_1,\ldots,t^*_N)$ is a collection of the center points based on partition $T=\cup^{N}_{i=1}[s_{i-1},s_i)$. It is not difficult to prove (see the details in Appendix B) that the conditional distribution of $\vess{\mu}{d}{N}$ given the data set $\mathcal{D}$ is a multivariate normal distribution with mean and covariance given by
\begin{eqnarray}\label{meanbar}
& & \ve{\bar{\mu}} \equiv \E(\vess{\mu}{d}{N}\mid \mathcal{D}) = \ve{\Psi}(\vesup{T}{*})\{\sigma^2\vesub{I}{n1}+(\vesub{I}{n1}-\vesub{\Sigma}{11})\vesub{\Gamma}{n1}\}^{-1} \{(\vesub{I}{n1}-\vesub{\Sigma}{11})\vesub{Y}{1} - \vesub{\Sigma}{12}\vesub{Y}{2}\}, \label{hlik2} \\
& & \ve{\Lambda} \equiv \Cov(\vess{\mu}{d}{N}\mid \mathcal{D}) = \vesub{\Gamma}{N}-\ve{\Psi}(\vesup{T}{*})\vess{\Gamma}{n1}{-1}\vesup{\Psi}{T}(\vesup{T}{*}) + \sigma^2 \ve{\Psi}(\vesup{T}{*})\vess{\Omega}{n1}{-1}\vesup{\Psi}{T}(\vesup{T}{*}),\nonumber
\end{eqnarray}
where $\ve{\Psi}(\vesup{T}{*})$ is the $N \times n_1$ covariance matrix between $\vess{\mu}{d}{N}$ and $\vesub{\tilde{\mu}}{d}$ with $(i,j)$th element $\gamma(t^*_i, t_j;\ve{\theta})$, $\vesub{\Gamma}{N}$ is the covariance matrix of $\vess{\mu}{d}{N}$ with $(i,j)$th element $\gamma(t^*_i,t^*_j;\ve{\theta})$, $\ve{\Sigma}$ is a $n\times n$ block matrix given by
\begin{equation*}
\ve{\Sigma}=\vesub{K}{n}(\vesub{K}{n}+\sigma^2\vesub{I}{n})^{-1} = \left(
                                     \begin{array}{cc}
                                       \vesub{\Sigma}{11} & \vesub{\Sigma}{12} \\
                                       \vesub{\Sigma}{21} & \vesub{\Sigma}{22} \\
                                     \end{array}
                                   \right)
\end{equation*}
with $\vesub{I}{n}$ being a $n\times n$ identity matrix, and $\vesub{\Omega}{n1}=\sigma^2\vesub{\Gamma}{n1} + \vesub{\Gamma}{n1}(\vesub{I}{n1}-\vesub{\Sigma}{11})\vesub{\Gamma}{n1}$. Therefore, to calculate $\widehat{\textup{mFDR}}_{\textup{I}}$ define in \eqref{mFDR1}, we draw $M$ samples $\{\hvess{\mu}{m}{N}: m=1,\ldots,M\}$ from the conditional distribution of $\vess{\mu}{d}{N}=(\mu_d(t^*_1),\ldots,\mu_d(t^*_N))^T$, where $\hvess{\mu}{m}{N}=(\widehat{\mu}^N_{m1},\ldots, \widehat{\mu}^N_{mN})^T$ is the $m$th $N$-dimensional sample predicting the values at time points $t^*_1,\ldots,t^*_N$. Then, we can approximate $\widehat{\textup{mFDR}}_{\textup{I}}$ by replacing $S_0(t^*_i)$ by its estimate $\widehat{S}_0(t^*_i)$. More specifically, note that
\begin{eqnarray*}
S_0(t^*_i) &=& {\rm P}(z(t^*_i)=0\mid \mathcal{D}) = \E[I\{|\mu_d(t^*_i)| \leq \Delta\}\mid \mathcal{D}]\\
 &=& \int I\{|\mu_d(t^*_i)| \leq \Delta\} \phi(\vess{\mu}{d}{N}\mid \ve{\bar{\mu}},\ve{\Lambda})d\vess{\mu}{d}{N}.
\end{eqnarray*}
Thus, $S_0(t^*_i)$ can be estimated by
\begin{equation*}
\widehat{S}_0(t^*_i)=\frac{1}{M}\sum^M_{m=1}I\{|\widehat{\mu}^N_{mi}| \leq \Delta\}.
\end{equation*}
Similarly, to implement procedure \eqref{mFDRtest}, we compute $S_1(t^*_i)$ and $S_2(t^*_i)$ by
\begin{eqnarray*}
\widehat{S}_1(t^*_i)&=&\frac{1}{M}\sum^M_{m=1}I\{ \widehat{\mu}^N_{mi} < -\Delta\}\\
\mbox{and}\ \widehat{S}_2(t^*_i)&=&\frac{1}{M}\sum^M_{m=1}I\{ \widehat{\mu}^N_{mi} > \Delta\},
\end{eqnarray*}
respectively.

{\rema The joint density function defined in \eqref{hlik} is the h-likelihood (Lee and Nelder, 1996) when we treat $\ve{\tilde{\mu}}$ and $\vesub{\tilde{\mu}}{d}$ as unobservable random variables. It contains all the information in the data for parameters $\ve{\Theta}$ and unobservable random variables $\ve{\tilde{\mu}}$ and $\vesub{\tilde{\mu}}{d}$ (Bj$\o$rnstad, 1996). The method discussed above can also be extended to a fully Bayesian way by assuming a hyper-prior distribution for $\ve{\Theta}$; see Shi and Choi (2011). }

{\rema Sun et al. (2015) used the similar approximation strategy to mimic the optimal procedure as in \eqref{mFDRtest}. But for implementation, they applied a Bayesian computational algorithm and drew MCMC samples during the iterations to estimate $S_0(t^*_i)$, which can be rather computationally intensive when the number of representative points $N$ is large. While for the proposed procedure, we can get the estimates of unknown parameters efficiently by using the nice proprieties of GPR models and estimate $S_k(t^*_i), k=0,1,2$ directly by generating the samples from multivariate normal distribution with mean and covariance given in \eqref{meanbar}. GPR models can cope with multiple covariates and thus the proposed method can be easily extend to problems in multivariate functional domain for example in 3-dimensional spatial domain or 4-dimensional temporal/spatial domain dynamical fMRI images.}

\section{Numerical study}\label{sec4}

\subsection{Simulation studies}

In this subsection, we conduct a set of simulation studies to assess the finite sample performance of the proposed method. The purpose is twofold. First, we compare our method with directional Benjamini-Hochberg procedure (Benjamini and Hochberg, 1995) and directional Benjamini-Yekutieli procedure (Benjamini and Yekutieli, 2001, 2005). Since both of them only work for discrete case where each hypothesis has its own observed data, in Example 1, we assume two curves observed at the same set of time points and restrict the analysis for testing hypotheses at this set to permit comparisons, which means we have $n_1=n_2=N$. Second, we evaluate the performance of our method in Example 2 to test hypotheses on a continuum $T$ with two curves observed at different discrete grid points.

{\sc {Example 1}} \quad We generate 200 datasets from model \eqref{model}, where both $\mu(\cdot)$ and $\mu_d(\cdot)$ are Gaussian processes with zero means and covariance functions $\kappa(t_i, t_j;\ve{\eta}) = 3\exp\{-(t_i-t_j)^2\}$ and $\gamma(t_i, t_j;\ve{\theta}) = 10\exp\left\{-\omega(t_i-t_j)^2/2\right\}$, respectively, implying two stationery processes, and the error processes $\epsilon_1(\cdot)$ and $\epsilon_2(\cdot)$ are white noise processes with zero mean and finite variance $\sigma^2=1$. For each simulated dataset, data are generated at $N=500$ time points $t_i \sim \textup{Uniform}([6,13])$. For all simulations, we choose $\Delta=0.80$ so that the expected proportion of time points with $|\mu_d(t)| \leq \Delta$ is 20$\%$ and set the nominal level as $\alpha=0.10$. To study the effects of correlation, we vary $\omega$ resulting in the curve $\mu_d(\cdot)$ from smooth to fluctuating.

\begin{figure}[htbp]
\centering \resizebox{12cm}{8cm}{\includegraphics{\patha 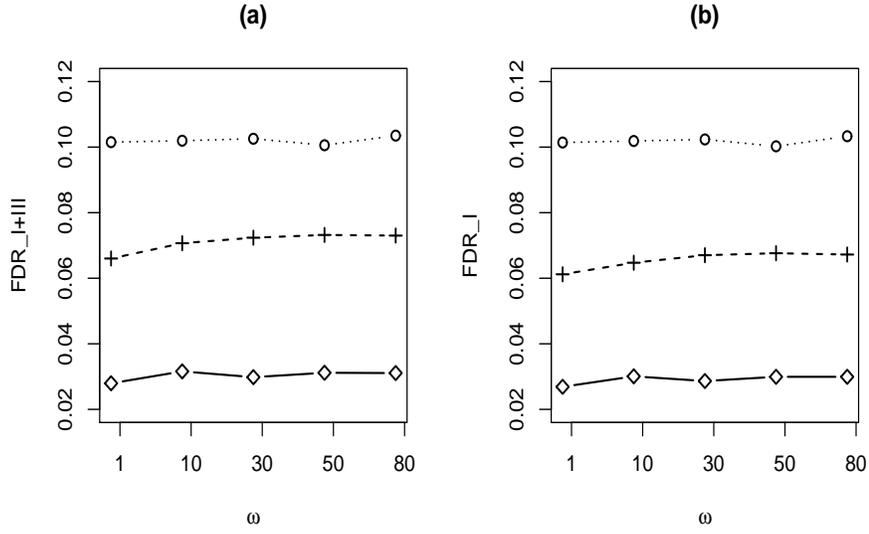}}
\caption{Comparison of directional Benjamini-Hochberg procedure ($+$), directional Benjamini-Yekutieli procedure ($\diamond$) and the proposed method ($o$): (a) FDR$_{\textup{I+III}}$ versus $\omega$; (b) FDR$_{\textup{I}}$ versus $\omega$.}\label{fig2}
\end{figure}

\begin{figure}[htbp]
\centering \resizebox{12cm}{8cm}{\includegraphics{\patha 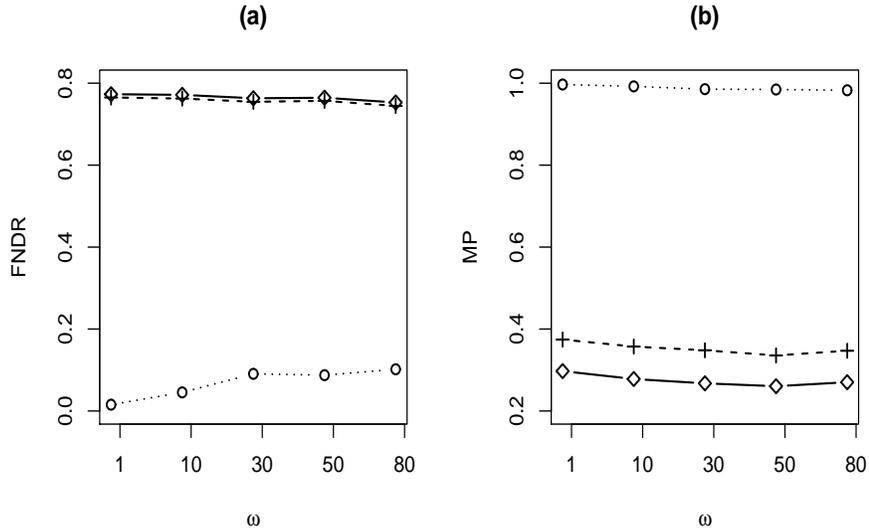}}
\caption{Comparison of directional Benjamini-Hochberg procedure ($+$), directional Benjamini-Yekutieli procedure ($\diamond$) and the proposed method ($o$): (a) FNDR versus $\omega$ under FDR$_{\textup{I+III}}$ at 0.10; (b) MP versus $\omega$ under FDR$_{\textup{I}}$ at 0.10.}\label{fig3}
\end{figure}

Figure~\ref{fig2} plots FDR$_{\textup{I+III}}$ and FDR$_{\textup{I}}$ as functions of $\omega$ at the nominal level 0.10 and Figure~\ref{fig3} shows the averages of FNDR and MP over the 200 datasets. We can see that the proposed method control FDR$_{\textup{I}}$ and FDR$_{\textup{I+III}}$ reasonably well. When $\omega$ becomes larger, there is a increasing chance to detect $\mu_d(t)$ to be non-null and declare it to be less than $-\Delta$ or larger than $\Delta$. That is, as the correlation of the signals decaying, it is more possible to make directional errors. As expected, Figure~\ref{fig3}(a) shows that the proposed method may have relatively large FNDR when $\omega$ is quite large. Correspondingly, Figure~\ref{fig3}(b) implies that it may encounter loss of power. Though the directional Benjamini-Yekutieli procedure accounts for dependence, it is the most conservative and therefore the least powerful. The directional Benjamini-Hochberg procedure, derived under the independence assumption, controls the FDR$_{\textup{I}}$ conservatively as the original Benjamini and Hochberg's procedure (Benjamini and Hochberg, 1995), which controls the FDR$_{\textup{I}}$ at a level smaller than the desired $\alpha$.

{\sc {Example 2}} \quad In this example, the true model is the same as in Example 1, except that the process $\mu(\cdot)$ is a Gaussian process with zero mean but a nonstationary covariance function $\kappa(t_i,t_j;\ve{\eta})=3\exp\{-(t_i-t_j)^2\}+3t_i t_j$. The sampling design for two curves is balanced ($n_1=n_2=200$), but irregular, and furthermore different across the two samples. Specifically, we assume that $\{t_{1i}:i=1, \ldots, n_1\}$ and $\{t_{2i}:i=1,\ldots,n_2\}$ are iid realizations from $\textup{Uniform}([6,13])$ with 20$\%$ overlapping. Predictions are made and tests of \eqref{test} are conducted at center points of $N=500$ equal-length subintervals covering the time range $[6,13]$. For all simulations, we set $\Delta=0.80$, $\alpha=0.10$ and vary the value of $\omega$ as in Example 1 and repeat our procedure 200 times for each configuration.

\begin{figure}[htbp]
\centering \resizebox{12cm}{8cm}{\includegraphics{\patha 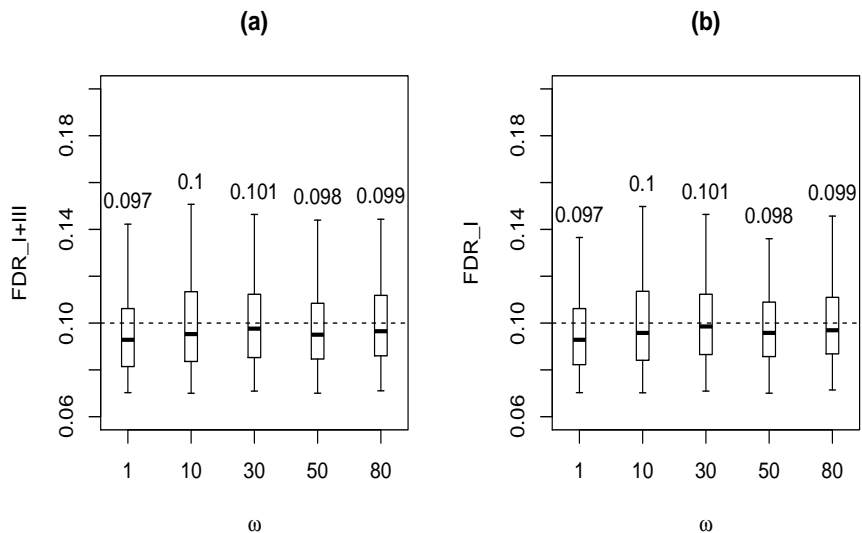}}
\caption{The boxplots of FDR$_{\textup{I+III}}$ and FDR$_{\textup{I}}$ based on 200 replications, respectively. The boxplots' horizontal lines are the 0.05, 0.25, 0.50, 0.75 and 0.95 quantiles of FDR$_{\textup{I+III}}$ and FDR$_{\textup{I}}$ versus $\omega$, and the numbers of above the boxplots are the means of FDR$_{\textup{I+III}}$ and FDR$_{\textup{I}}$.}\label{fig4}
\end{figure}

\begin{figure}[htbp]
\centering \resizebox{12cm}{8cm}{\includegraphics{\patha 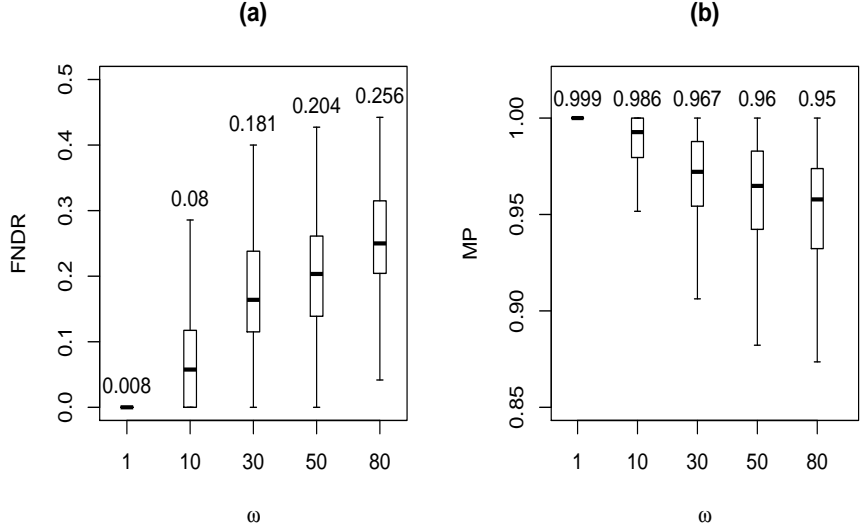}}
\caption{The boxplots of FNDR and MP based on 200 replications, respectively. The boxplots' horizontal lines are the 0.05, 0.25, 0.50, 0.75 and 0.95 quantiles of FNDR and MP versus $\omega$, and the numbers of above the boxplots are the means of FNDR and MP.}\label{fig5}
\end{figure}

Figure~\ref{fig4} depicts the distribution of FDR$_{\textup{I+III}}$ and FDR$_{\textup{I}}$ and Figure~\ref{fig5} presents the distribution of FNDR and MP over 200 replications. We can see that the proposed method maintains FDR$_{\textup{I+III}}$ and FDR$_{\textup{I}}$ properly no matter the curve $\mu_d(\cdot)$ is smooth or wiggly. It is in accordance with Theorem~\ref{theo2} in Section~\ref{sec3-2}. And it implies that the proposed procedure is robust under different magnitudes of dependence across the values of $\mu_d(\cdot)$. Moreover, the boxplots of FNDR and MP show that the proposed procedure is powerful, where the MP is 0.95 even when $\omega$ is very large.

To investigate the consistency of the estimated directional errors, the values of the estimated $\E\{\mathcal{L}(N_1)\}$, $\E\{\mathcal{L}(N_2)\}$, $\E\{\mathcal{L}(N_3)\}$, $\E\{\mathcal{L}(R_0)\}$, $\E\{\mathcal{L}(R_1)\}$ and $\E\{\mathcal{L}(R_2)\}$ are averaged so that the directional errors are calculated and regarded as the true values. Table~\ref{tab2} compares them with the estimated directional errors when $\omega=80$. We observe that the proposed procedure gives consistent estimators. Slight underestimation of mFDR explains slightly liberal control of directional FDRs when $\omega$ is large.

\begin{center}{\small
\begin{table}[htb!]
\caption{\label{tab2} {True errors and averages (standard deviation) of estimated errors when $\omega=80$} \hspace{6.5cm}}{\small {\footnotesize \tabcolsep
0.9cm
\renewcommand{\arraystretch}{1.25}
\begin{tabular}{c|cccc}
\hline
\multicolumn{5}{c}{When controlling FDR$_{\textup{I}}$ at 0.1}\\
\hline
Errors & mFDR$_{\textup{I}}$ & mFDR$_{\textup{III}}$ & mFDR$_{\textup{I+III}}$ &  mFNDR\\
\hline
True & 0.099 &  0.003 & 0.102 & 0.260\\
Estimated &  0.098 & 0.003 & 0.101 & 0.259\\
 & (0.002) & (0.002) & (0.002) & (0.040)\\
\hline
\multicolumn{5}{c}{When controlling FDR$_{\textup{I+III}}$ at 0.1}\\
\hline
Errors & mFDR$_{\textup{I}}$ & mFDR$_{\textup{III}}$ & mFDR$_{\textup{I+III}}$ &  mFNDR\\
\hline
True & 0.096 &  0.003 & 0.099 & 0.256\\
Estimated &  0.095 & 0.003 & 0.098 & 0.254\\
 & (0.002) & (0.002) & (0.000) & (0.040)\\
\hline
\end{tabular}%
}}
\end{table}
}
\end{center}

\subsection{Real data analysis}

To illustrate the proposed method, we analyze BLC mean correct latency for action video game players (AVGPs) and non action video game players (NAVGPs). The data consists of 84 girls and 57 boys from primary and secondary schools, aging from 6 to 13 years old. They were recruited to answer the video game playing questionnaire. Using data from the questionnaire, which were collected separately from children and from their parents for verification, these 141 students were subdivided into two groups: the AVGPs group (56$\%$) and the NAVGPs group (44$\%$). Then, they were required to finish the Big/Little Circle test via an action video game, which was defined as a video game genre that emphasizes hand-eye coordination and reaction-time. Our objective is to detect the areas of age that the significant differences between AVGPs group and NAVGPs group occur.

We use the Gaussian process regression model \eqref{model} to fit the data for each group. The estimated mean curves corresponding to AVGPs group and NAVGPs group are given in Figure~\ref{fig6}. {We can see that there are some crossings between these two curves. We first consider a test with from \eqref{test} and $\Delta=20$, chosen by our collaborators in neuroscience.

We generate samples based on the conditional distribution of $\vess{\mu}{d}{N}$ on center points of 500 equal-length subintervals covering the age range $[6,13]$, and test the hypotheses at each time point. Figure~\ref{fig6} shows the significant and non-significant areas detected by the proposed procedure, when controlling FDR$_{\textup{I+III}}$ at level 0.10. Aging from 6 to 9, the NAVGPs have significantly higher BLC mean correct latency than AVGPs, while after 9 years old, they have non-significant differences. It implies that the video game-based therapy may have significant effect on children with hemiplegia aging from 6 to 9 years old, while it may have limited help with of some of symptoms when they are more than 9 years old. The proposed procedure reports $\widehat{\textup{mFDR}_{\textup{I}}}=0.08$ and $\widehat{\textup{mFDR}_{\textup{III}}}=0.02$, indicating the Type III errors account for about 20$\%$ of mFDR$_{\textup{I+III}}$. It reports $\widehat{\textup{mFNDR}}=0.22$, implying that the means of BLC mean correct latency of NAVGPs and AVGPs groups could have differences larger than $\Delta=20$ in 22$\%$ of areas of age after 9 years old.

\begin{figure}[htbp]
\centering \resizebox{12cm}{8cm}{\includegraphics{\patha 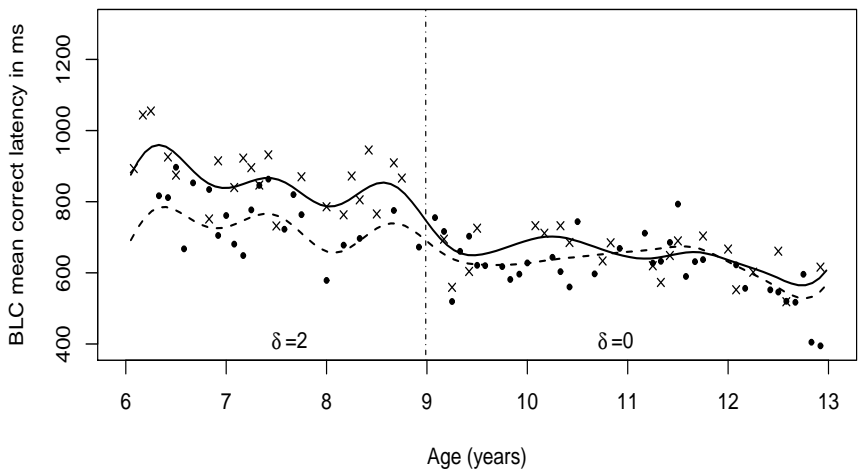}}
\caption{The significant and non-significant areas detected by the proposed procedure under mFDR$_{\textup{I+III}}$ control at 0.10. Aging from 6 to 9, the NAVGPs have significantly higher BLC mean correct latency than AVGPs ($\delta(t)=2$), while after 9 years old, they have non-significant differences ($\delta(t)=0$). The solid and the dash lines represent the estimated mean curves for the NAVGPs group ($\times$) and the AVGPs group ($\centerdot$), respectively. The estimates of errors are: $\widehat{\textup{mFDR}_{\textup{I}}}=0.08$, $\widehat{\textup{mFDR}_{\textup{III}}}=0.02$, and $\widehat{\textup{mFNDR}}=0.22$.}\label{fig6}
\end{figure}

Using different values of $\Delta$ in \eqref{test} makes the method very flexible. Figure~\ref{fig7} presents the results with $\Delta=1$ and $\Delta =100$. The estimated mFDR$_\textup{I}$, mFDR$_\textup{III}$ and mFNDR are also calculated and presented. The former indicates there are two significant areas: one from age 6 to 9.2 and the other from 9.6 to 10.6. Consequently, with a smaller $\Delta$ mFNDR increases, i.e. there could exist 44$\%$ of areas, among declared non-significant areas, that the mean difference of these two groups is larger than $\Delta=1$. The results for $\Delta =100$ imply that there is no detected significant area, while there would exist $30\%$ of areas that the mean difference of two groups is larger than $\Delta=100$ in whole age range [6,13]. It is not surprising that there is no rejection at all when $\Delta\geq 100$. This rather large number makes the result meaningless. In general, the choice of $\Delta$ depends on a scientific question of interest.

\begin{figure}[htbp]
\centering \resizebox{7cm}{6cm}{\includegraphics{\patha 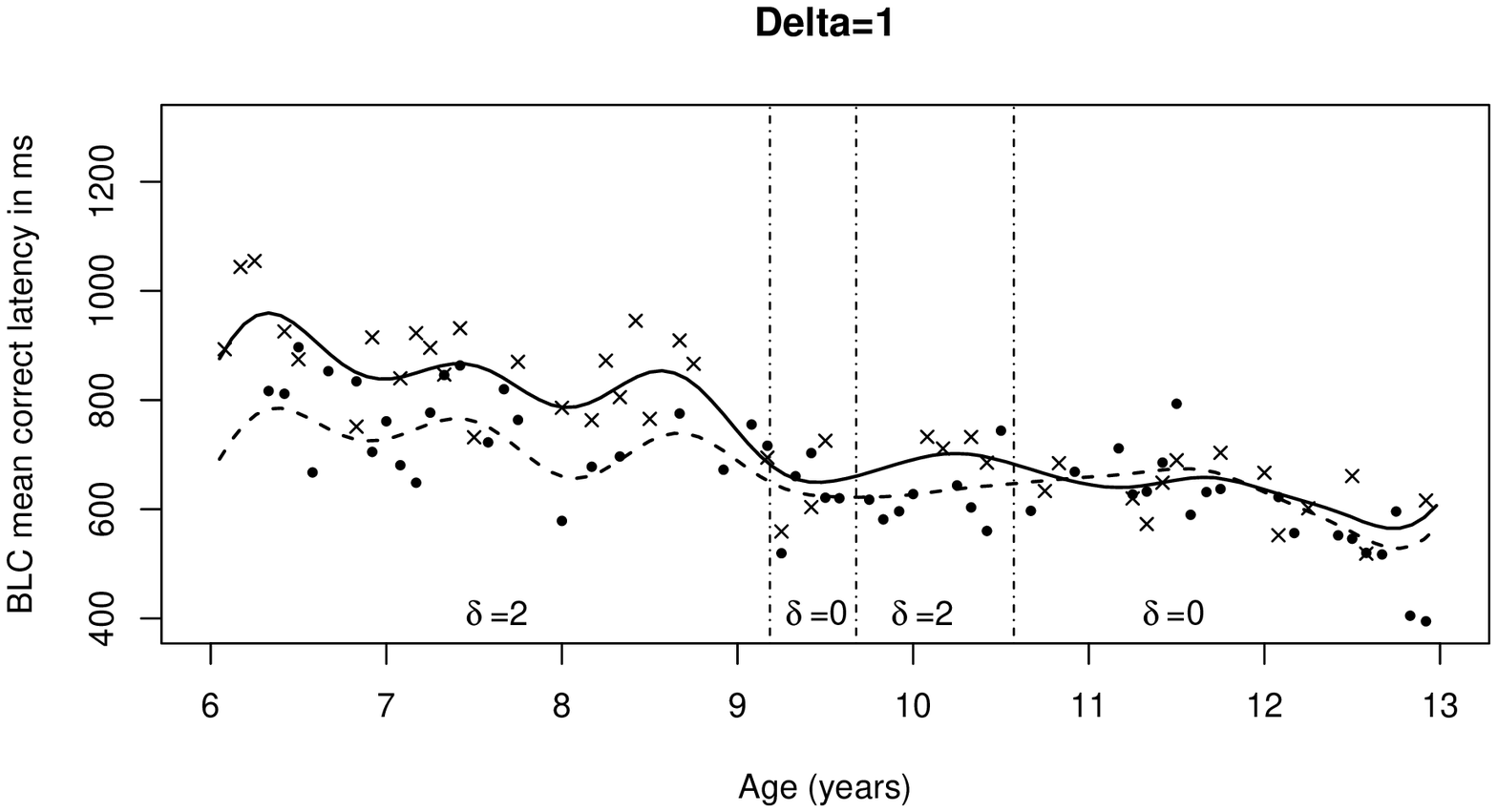}} \
\resizebox{7cm}{6cm}{\includegraphics[width=0.45\linewidth]{\patha 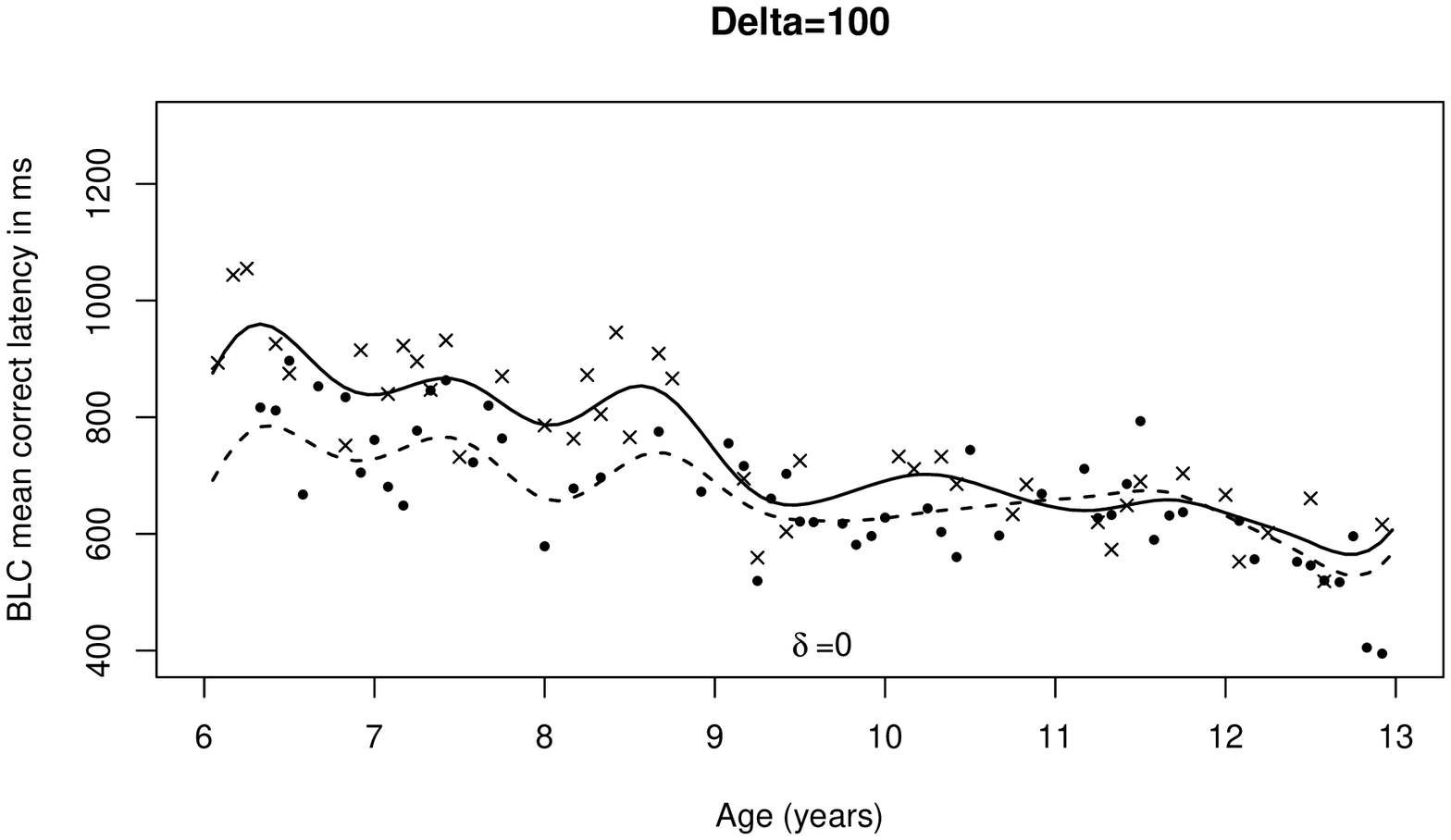}}
\caption{The significant and non-significant areas detected by the proposed procedure under mFDR$_{\textup{I+III}}$ control at 0.10. Left ($\Delta=1$): aging from 6 to 9.2 and from 9.6 to 10.6, the NAVGPs have significantly higher BLC mean correct latency than AVGPs ($\delta(t)=2$), while they have non-significant differences at other ages ($\delta(t)=0$). The estimates of errors are: $\widehat{\textup{mFDR}_{\textup{I}}}=0.01$, $\widehat{\textup{mFDR}_{\textup{III}}}=0.09$, and $\widehat{\textup{mFNDR}}=0.44$; right ($\Delta=100$): there is no rejection which implies that no significant area is detected. The estimates of errors are: $\widehat{\textup{mFDR}_{\textup{I}}}=0$, $\widehat{\textup{mFDR}_{\textup{III}}}=0$, and $\widehat{\textup{mFNDR}}=0.30$.} \label{fig7}
\end{figure}

\section{Equivalence tests} \label{sec7}

A statistical hypothesis test is a decision rule to check whether the null hypothesis is justifiable given the observed data. We could reject the null hypothesis when there is strong evidence that it is wrong, but we could never prove it. Therefore, failure to reject $H_0(t)$ in \eqref{test} does not mean that the difference between mean functions of two curves $Y_1(t)$ and $Y_2(t)$ is no more than $\Delta$ at time $t$. To demonstrate similarity rather than showing differences, we sometimes need to put the similarity hypothesis into the alternative. We might thus consider the multiple testing
\begin{eqnarray}\label{test2}
& & H_{01}(t):\mu_d(t) < -\Delta^{\textup{E}}\quad \mbox{ or }\quad H_{02}(t):  \mu_d(t) > \Delta^{\textup{E}} \nonumber\\
&\mbox{versus}& H_1(t): |\mu_d(t)| \leq \Delta^{\textup{E}},
\end{eqnarray}
where $\Delta^{\textup{E}}$ is called equivalence margin that is typically chosen as a limit below which differences are practically meaningful, and we call the test \eqref{test2} the equivalence testing for functional data.

Equivalence tests have gained increasing attention during the past two decades. The goal of an equivalence test is to establish practical equivalence, which is popular used in application areas such as medicine and biology. There are lots of procedures that have been proposed to conduct equivalence tests for scalar data. For example, Schuirmann (1987) proposed the two one-sided tests procedure for bioequivalence; Anderson and Hauck (1990) suggested the comparison of both mean and variance of the two responses when assess a generic drug's performance relative to a brand name drug; Brown et al. (1997) developed an unbiased test for the bioequivalence problem; Wang et al. (1999) discussed ways to construct a test simultaneously for all the individual pharmacokinetic parameters; Romano (2005) proposed a optimal test for testing the mean of a multivariate normal mean. Other relevant works include Chow and Liu (1992), Berger and Hsu (1996), Meyners (2012) and some of the references therein. However, in some cases the question of practical equivalence cannot be reduced to a hypothesis regarding scalar data. Recently, Fogarty and Small (2014) extended the equivalence testing framework to the functional regime. They considered an equivalence testing for overall mean difference. But they cannot test areas of the function domain with location parity. Therefore, it will be interesting to extend the proposed idea to equivalence testing \eqref{test2}.

Let $z^{\textup{E}}(t)$ be the underlying state at time $t$. We set $z^{\textup{E}}(t)=1$ or 2 if hypothesis at time $t$ is the null 1 or 2 and $z^{\textup{E}}(t)=3$ if hypothesis at time $t$ is the alternative. Let $\delta^{\textup{E}}(t) \in \{1,2,3\}$ be a decision rule for the hypothesis \eqref{test2}. Let $R^{\textup{E}}_k=\{t\in T: \delta^{\textup{E}}(t)=k\}$ and $V^{\textup{E}}_{jk}=\{t \in T: z^{\textup{E}}(t)=j, \delta^{\textup{E}}(t)=k\}$ for $j,k=1,2,3$. Similar to directional two-sided test \eqref{test}, there also exist three types of errors in equivalence testing \eqref{test2}. Table~\ref{tab3} sums up the possible outcomes of multiple testing with two nulls. Then, $\mathcal{L}(N^{\textup{E}}_1) = \mathcal{L}(V^{\textup{E}}_{13}) + \mathcal{L}(V^{\textup{E}}_{23})$, $\mathcal{L}(N^{\textup{E}}_2)=\mathcal{L}(V^{\textup{E}}_{31})+\mathcal{L}(V^{\textup{E}}_{32})$ and $\mathcal{L}(N^{\textup{E}}_3)=\mathcal{L}(V^{\textup{E}}_{12})+\mathcal{L}(V^{\textup{E}}_{21})$ are the sizes of areas corresponding to Type I, Type II and Type III errors, respectively, where $\mathcal{L}(\cdot)$ is the Lebesgue measure on $T$. Hence, we define the marginal false discovery rate as $\textup{mFDR}^{\textup{E}}  = \E\{\mathcal{L}(N^{\textup{E}}_1)\} / \E\{\mathcal{L}(R^{\textup{E}}_3)\}$, the marginal false nondiscoveary rate for Type II error as $\textup{mFNDR}^{\textup{E}}_{\textup{II}}  =  \E\{\mathcal{L}(N^{\textup{E}}_2)\} / \E\{\mathcal{L}(R^{\textup{E}}_1\cup R^{\textup{E}}_2)\}$ and that for Type III error as $\textup{mFNDR}^{\textup{E}}_{\textup{III}}  = \E\{\mathcal{L}(N^{\textup{E}}_3)\} / \E\{\mathcal{L}(R^{\textup{E}}_1\cup R^{\textup{E}}_2)\}$. And for simplicity, let $\textup{mFNDR}^{\textup{E}}_{\textup{II+III}} = \textup{mFNDR}^{\textup{E}}_{\textup{II}} + \textup{mFNDR}^{\textup{E}}_{\textup{III}}$.

\begin{center}{\small
\begin{table}[htb!]
\caption{\label{tab3} {Outcomes of multiple testing with two nulls} \hspace{6.5cm}}{\small {\footnotesize \tabcolsep
0.2cm
\renewcommand{\arraystretch}{1.2}
\begin{tabular}{l|ccc|c}
\hline
& Declared as null 1 & Declared as null 2 & Declared as alternative & Total\\
& $\delta^{\textup{E}}(t)=1$ & $\delta^{\textup{E}}(t)=2$ & $\delta^{\textup{E}}(t)=3$ & \\ \hline
Null 1 ($z^{\textup{E}}(t)=1$)& $V^{\textup{E}}_{11}$     &  $V^{\textup{E}}_{12}$ (\emph{Type III error}) & $V_{13}$ (\emph{Type I error}) & $T^{\textup{E}}_1$\\
Null 2 ($z^{\textup{E}}(t)=2$) &   $V^{\textup{E}}_{21}$ (\emph{Type III error}) & $V^{\textup{E}}_{22}$ & $V^{\textup{E}}_{23}$ (\emph{Type I error}) & $T^{\textup{E}}_2$\\
Alternative ($z^{\textup{E}}(t)=3$) &   $V^{\textup{E}}_{31}$ (\emph{Type II error}) & $V^{\textup{E}}_{32}$ (\emph{Type II error}) & $V^{\textup{E}}_{33}$ & $T^{\textup{E}}_3$\\
\hline
Total & $R^{\textup{E}}_1$ & $R^{\textup{E}}_2$  & $R^{\textup{E}}_3$ & $T$\\
\hline
\end{tabular}%
}}
\end{table}
}
\end{center}

We applied the equivalence testing \eqref{test2} to the executive function study. The results are presented in Figure~\ref{fig8}. The non-significant areas (i.e. the mean curves are different) obtained by using $\Delta^{\textup{E}}=140$ is similar to the ones using test \eqref{test} with $\Delta=20$ (see Figure~\ref{fig6}). One reason might be the $\textup{mFDR}^{\textup{E}}$ (analogous to Type I error) controlled here is actually the mFNDR (analogous to Type II error) in the multiple testing \eqref{test}, and the $\textup{mFNDR}^{\textup{E}}_{\textup{II+III}}$ (analogous to the sum of Type II and III errors) minimized in the equivalence testing \eqref{test2} is actually the mFDR$_{\textup{I+III}}$ (analogous to the sum of Type I and III errors) in test \eqref{test}, which shows the clear differences between these two different types of test. As expected when $\Delta^{\textup{E}}\leq 100$, there is no rejection.

\begin{figure}[htbp]
\centering \resizebox{7cm}{6cm}{\includegraphics{\patha 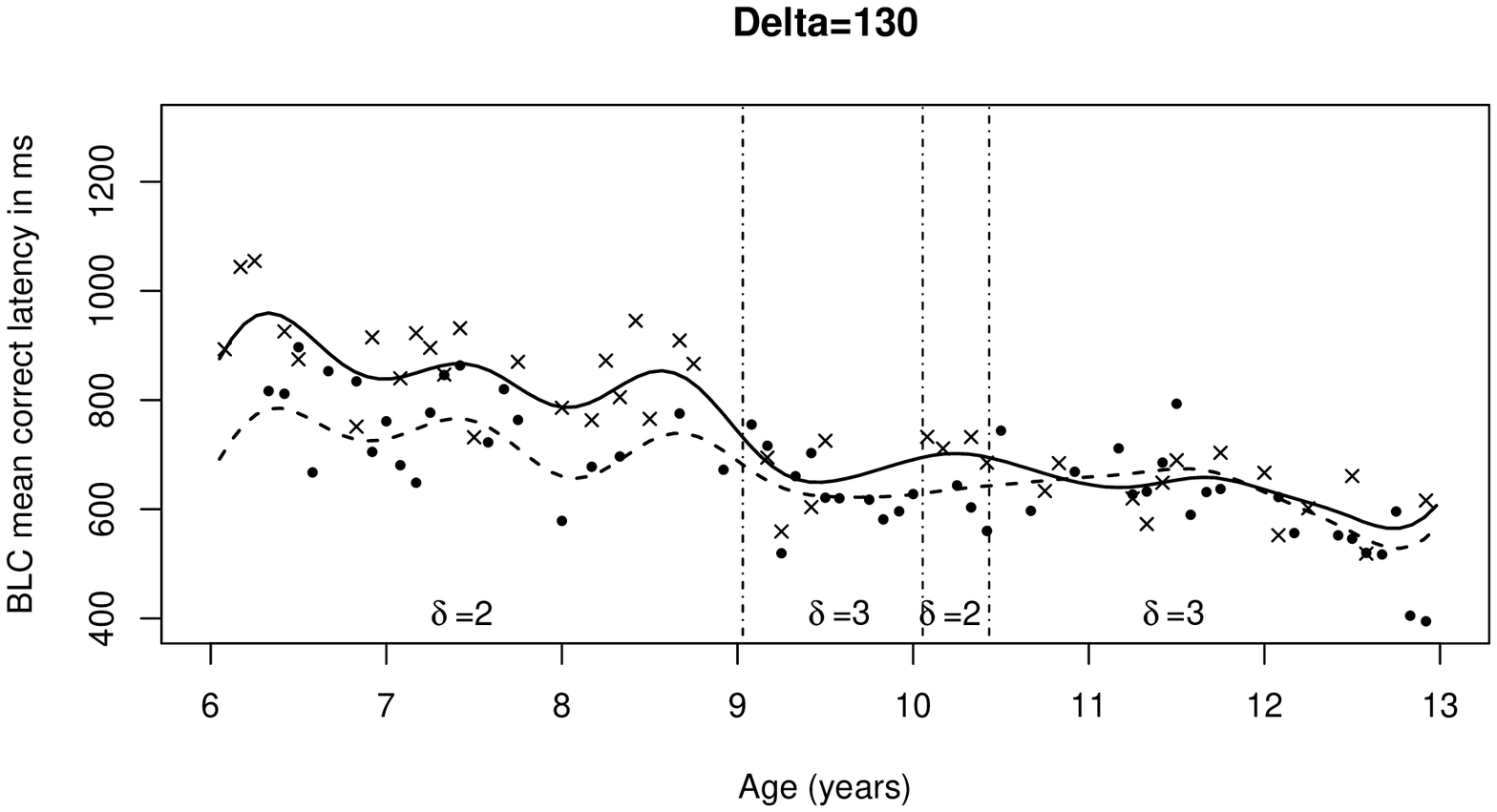}}
\resizebox{7cm}{6cm}{\includegraphics{\patha 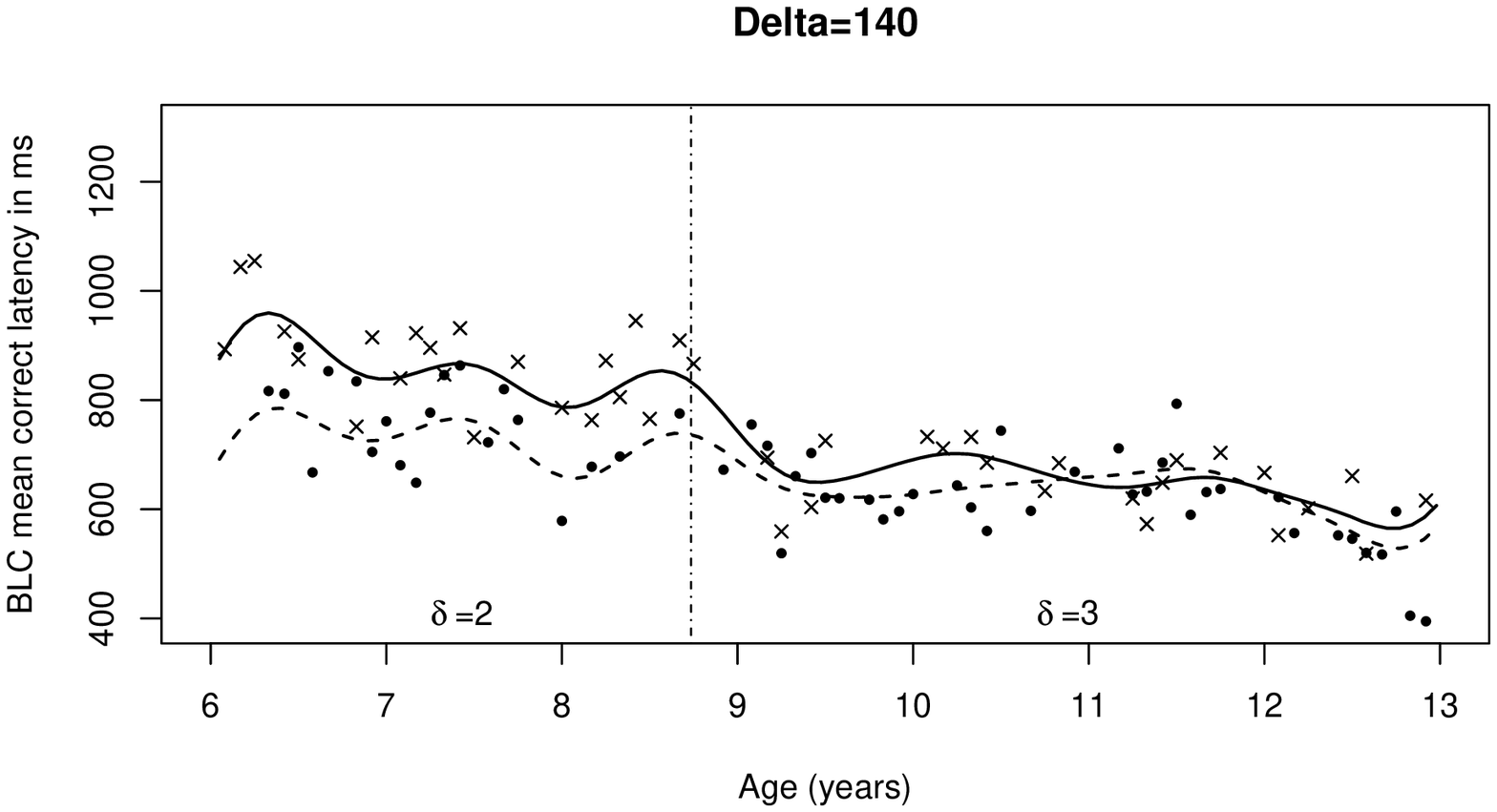}}
\caption{The equivalent and non-equivalent areas detected by the proposed procedure under $\textup{mFDR}^{\textup{E}}$ control at 0.10. Left ($\Delta^{\textup{E}}=130$): aging from 9 to 10 and after 10.5, the BLC mean correct latency of NAVGPs and AVGPs are similar ($\delta^{\textup{E}}(t)=3$), while the NAVGPs have higher BLC mean correct latency than AVGPs at other ages ($\delta^{\textup{E}}(t)=2$). The estimates of errors are:
$\widehat{\textup{mFDR}^{\textup{E}}}=0.10$, $\widehat{\textup{mFNDR}^{\textup{E}}_{\textup{II}}}=0.25$, and $\widehat{\textup{mFNDR}^{\textup{E}}_{\textup{III}}}=0.001$; Right ($\Delta^{\textup{E}}=140$): aging after 8.8, the BLC mean correct latency of NAVGPs and AVGPs are similar ($\delta^{\textup{E}}(t)=3$), while the NAVGPs have higher BLC mean correct latency than AVGPs from 6 to 8.8 ($\delta^{\textup{E}}(t)=2$). The estimates of errors are: $\widehat{\textup{mFDR}^{\textup{E}}}=0.10$, $\widehat{\textup{mFNDR}^{\textup{E}}_{\textup{II}}}=0.27$, and $\widehat{\textup{mFNDR}^{\textup{E}}_{\textup{III}}}=0$. \label{fig8}}
\end{figure}

\section{Discussion}\label{sec5}

In this paper we proposed a method based on large scale multiple testing to detect differences of the means of two curves. It can automatically detect the significant areas  and at the same time control the directional error.  By taking advantage of the functional nature of the data, we introduce a nonparametric  Gaussian process regression model for simultaneous two-sided tests. We are thus able to make inference at any point in a continuum and derive a procedure which optimally controls directional false discovery rates. To make it workable in practice, an approximation procedure is proposed via a finite approximation strategy. We show that the proposed procedure controls directional false discovery rates at any specified level asymptotically.

Related to the topic discussed in this paper, some interesting problems are worth further development. Though simulation studies validate the good control ability of the proposed procedure over both Type I and directional errors, the estimation of the unknown model parameters may affect the power of the testing method. It is therefore important for us to discuss the asymptotic optimality of the data-driven procedure with estimated model parameters in a more systematic fashion. And this paper focuses mainly on the problem defined in one-dimensional domain. It will be interesting to extend the idea to more complicated case, such as the problem defined in two- or three-dimensional spatial domain, or in temporal-spatio domain. Gaussian process regression model can cope with problems with multidimensional covariates. This good feature makes such extension feasible. On the other side of the spectrum, Fogarty and Small (2014) considered an equivalence testing for overall mean difference with dynamic bands. To extend the equivalence testing \eqref{test2} to a more general case with dynamic lower and upper equivalence bands is another interesting direction for future investigation.

\section*{Acknowledgements}

We would like to thank Professor J. Eyre of the Institute of Neuroscience of Newcastle University in UK for allowing us to use their experimental data. Xu was supported by the Natural Science Foundation of Jiangsu Province, China (No. BK20140617). Lee was supported by the Brain Research Program through the National Research Foundation of Korea (NRF) funded by the Ministry of Science, ICT and Future Planning (2014M3C7A1062896).

\newpage
\section*{Appendix}

\subsection*{Appendix A. Technical Proofs}

{\bf{Proof of Theorem~\ref{theo1}}}. We first prove Theorem~\ref{theo1}(1). If $\lambda_2=1$ and $\lambda_1=\lambda_3=\lambda$, the loss function \eqref{loss} becomes
\begin{equation*}
L(\delta,z;\lambda)=\mathcal{L}(N_2) + \lambda\{\mathcal{L}(N_1)+\mathcal{L}(N_3)\},
\end{equation*}
which can be re-written as
\begin{eqnarray*}
L(\delta,z;\lambda) &=& \sum^2_{k=1}\int_T I(z(t)=k)I(\delta(t)=0)d\mathcal{L}(t) + \lambda\left\{\sum^2_{k=1}\int_T I(z(t)=0)I(\delta(t)=k)d\mathcal{L}(t)\right.\\
& & \left. + \sum^2_{j=1}\sum_{k\neq 0,j} \int_T I(z(t)=k)I(\delta(t)=j)d\mathcal{L}(t)\right\}.
\end{eqnarray*}
Then, the posterior classification risk is
\begin{eqnarray*}
\E\{L(\delta,z;\lambda)\mid \mathcal{D}\}  &=& \sum^2_{k=1}\int_T I(\delta(t)=0) {\rm P}(z(t)=k\mid \mathcal{D})d\mathcal{L}(t) + \lambda\left\{\sum^2_{k=1}\int_T I(\delta(t)=k)\right.\\
& & \left.{\rm P}(z(t)=0\mid \mathcal{D})d\mathcal{L}(t)
+ \sum^2_{j=1}\sum_{k\neq 0,j} \int_T I(\delta(t)=j) {\rm P}(z(t)=k\mid \mathcal{D})d\mathcal{L}(t)\right\}\\
&=& \int_T \left\{I(\delta(t)=0) {\rm P}(z(t)\neq 0\mid \mathcal{D}) + \lambda \sum^2_{k=1}I(\delta(t)=k){\rm P}(z(t)\neq k\mid \mathcal{D})\right\}d\mathcal{L}(t)\\
&=& \int_T {\rm P}(z(s)\neq 0\mid \mathcal{D}) \left\{I(\delta(t)=0) + \sum^2_{k=1}I(\delta(t)=k)\frac{\lambda {\rm P}(z(t)\neq k\mid \mathcal{D})}{{\rm P}(z(s)\neq 0\mid \mathcal{D})}\right\}d\mathcal{L}(t).
\end{eqnarray*}
Therefore, the optimal decision rule  $\delta^{(I+III)}=\{\delta^{(I+III)}(t): t\in T\} = \textup{argmin}_{\delta} \E\{L(\delta,z;\lambda)|\mathcal{D}\}$ is
\begin{eqnarray*}
\delta^{(I+III)}(t) &=& k\ \mbox{if}\ \frac{{\rm P}(z(t)\neq 0\mid \mathcal{D})}{{\rm P}(z(t)\neq k\mid \mathcal{D})} > \lambda\ \mbox{and}\ {\rm P}(z(t)=k\mid \mathcal{D}) = \max_{j=1,2} {\rm P}(z(t)=j\mid \mathcal{D}),\\
&=& 0\ \mbox{otherwise},
\end{eqnarray*}
which finishes the proof of Theorem~\ref{theo1}(1). Similar arguments can be used to prove Theorem~\ref{theo1}(2).\hfill $\fbox{}$

\
{\bf{Proof of Theorem~\ref{theo3}}}. Given an mFDR$_{\textup{I+III}}$ level $\alpha$, consider a decision rule $\delta=\{\delta(t): t\in T\}$ with mFDR$_{\textup{I+III}}\{\delta\} \leq \alpha$. Let $R$ be the expected rejection area for $\delta$. Define $\Upsilon(t) = \{\min_{1\leq k \leq 2}{\rm P}(z(t)=k\mid \mathcal{D}) + {\rm P}(z(t)= 0\mid \mathcal{D})\} / {\rm P}(z(t)\neq 0\mid \mathcal{D})$. Then, according to the definition of $\delta^{(I+III)}$, its corresponding expected rejection area is
\begin{equation*}
R(\lambda)=\E \int_T I(\Upsilon(t)\leq \lambda^{-1})d\mathcal{L}(t)= \int_T {\rm P}(\Upsilon(t)\leq \lambda^{-1})d\mathcal{L}(t).
\end{equation*}
Hence, $R(\lambda)$ is decreasing with $\lambda$. In addition, it is easy to see that
\begin{equation*}
\lim_{\lambda \rightarrow 0}\frac{R(\lambda)}{\mathcal{L}(T)}=1,\ \mbox{and}\ \lim_{\lambda \rightarrow \infty} R(\lambda)=0.
\end{equation*}
Consequently, for a given expected rejection area $R$ determined by $\delta$, there exists a unique $\lambda(R)$ such that the decision rule $\delta^{(I+III)}$ has the same expected rejection area.

Further, for $\delta^{(I+III)}$, define TD$_{\delta^{(I+III)}}$, FD$_{\delta^{(I+III)} I}$ and FD$_{\delta^{(I+III)} III}$ as the expected true discovery area, expected false discovery area related to Type I error and expected false discovery area related to Type III error, respectively. Then, we have
\begin{eqnarray*}
\textup{TD}_{\delta^{(I+III)}} &=& \sum^2_{k=1}\E \int_T I(z(t)=k)I(\delta^{(I+III)}(t)=k)d\mathcal{L}(t),\\
\textup{FD}_{\delta^{(I+III)} I} &=& \sum^2_{k=1}\E \int_T I(z(t)=0)I(\delta^{(I+III)}(t)=k)d\mathcal{L}(t),\\
\textup{FD}_{\delta^{(I+III)} III} &=& \sum^2_{j=1}\sum_{k\neq 0, j} \E \int_T I(z(t)=k)I(\delta^{(I+III)}(t)=j)d\mathcal{L}(t),
\end{eqnarray*}
and $R(\lambda)=\textup{TD}_{\delta^{(I+III)}} +\textup{FD}_{\delta^{(I+III)} I} + \textup{FD}_{\delta^{(I+III)} III}$. Similarly, let TD$_{\delta}$, FD$_{\delta I}$ and FD$_{\delta III}$ be the expected true discovery area, expected false discovery area related to Type I error and expected false discovery area related to Type III error for $\delta$, respectively. Then, it also holds that $R(\lambda)=\textup{TD}_{\delta} +\textup{FD}_{\delta I}+\textup{FD}_{\delta III}$. For $\zeta=\delta^{(I+III)}, \delta$, consider the loss function
\begin{eqnarray*}
L(z,\zeta)&=&\mathcal{L}(N_2) + \lambda\{\mathcal{L}(N_1)+\mathcal{L}(N_3)\}\\
&=& \sum^2_{k=1}\int_T I(z(t)=k)I(\zeta(t)=0)d\mathcal{L}(t) + \lambda\left\{\sum^2_{k=1}\int_T I(z(t)=0)I(\zeta(t)=k)d\mathcal{L}(t)\right.\\
& & \left. + \sum^2_{j=1}\sum_{k\neq 0,j} \int_T I(z(t)=k)I(\zeta(t)=j)d\mathcal{L}(t)\right\}.
\end{eqnarray*}
Then, the risk for $\delta$ and $\delta^{(I+III)}$ is
\begin{eqnarray*}
\E L(z,\zeta)  &=& \sum^2_{k=1}\E \int_T I(z(t)=k)\{1-I(\zeta(t)\neq 0)\}d\mathcal{L}(t) + \lambda(\textup{FD}_{\zeta I}+\textup{FD}_{\zeta III})\\
&=& \int_T\sum^2_{k=1}{\rm P}(z(t)=k)d\mathcal{L}(t) - \E \int_T \sum^2_{k=1}I(z(t)=k)I(\zeta(t)=k)d\mathcal{L}(t)\\
& & - \E \int_T \sum^2_{j=1}\sum_{k\neq 0, j} I(z(t)=k)I(\zeta(t)=j)d\mathcal{L}(t) + \lambda(\textup{FD}_{\zeta I}+\textup{FD}_{\zeta III}) \\
&=& \int_T\sum^2_{k=1}{\rm P}(z(t)=k)d\mathcal{L}(t) + \lambda(\textup{FD}_{\zeta I}+\textup{FD}_{\zeta III}) - (\textup{TD}_{\zeta}+\textup{FD}_{\zeta III}).
\end{eqnarray*}
Since $\E L(z,\delta^{(I+III)}) \leq \E L(z, \delta)$, it implies that
$\textup{FD}_{\delta^{(I+III)} I}+\textup{FD}_{\delta^{(I+III)} III} \leq \textup{FD}_{\delta I}+\textup{FD}_{\delta III}$
and $\textup{TD}_{\delta^{(I+III)}}+\textup{FD}_{\delta^{(I+III)} III} \geq \textup{TD}_{\delta}+\textup{FD}_{\delta III}$.
Therefore,
\begin{equation*}
\textup{mFDR}_{\textup{I+III}}\{\delta^{(I+III)}\} = \frac{\textup{FD}_{\delta^{(I+III)} I}+\textup{FD}_{\delta^{(I+III)} III}}{R(\lambda)}\leq \frac{\textup{FD}_{\delta I}+\textup{FD}_{\delta III}}{R(\lambda)}=\textup{mFDR}_{\textup{I+III}}\{\delta\}\leq \alpha,
\end{equation*}
and
\begin{equation*}
\textup{mFNDR}\{\delta^{(I+III)}\} = \frac{\textup{TD}_{\delta^{(I+III)}}+\textup{FD}_{\delta^{(I+III)} III}}{\mathcal{L}(T)-R(\lambda)}\leq \frac{\textup{TD}_{\delta}+\textup{FD}_{\delta III}}{\mathcal{L}(T)-R(\lambda)} \leq \textup{mFNDR}\{\delta\}.
\end{equation*} \hfill $\fbox{}$

\
To prove the procedure \eqref{mFDRtest} is asymptotically valid for FDR$_{\textup{I}}$ control, we first need the following regularity conditions.
\begin{itemize}
\item[C1] Let $\rho>0$ be a small positive constant. For $\mu_0=-\Delta$ or $\Delta$, $\int_T {\rm P}(|\mu_d(t)-\mu_0| < \rho)d\mathcal{L}(t) \rightarrow 0$ as $\rho \rightarrow 0$.
\item[C2] Let $\mu^N_d(t)=\sum^N_{i=1}\mu_d(t^*_i)I(s_{i-1}\leq t < s_i)$. Assume the sequence of partitions $\{\cup^N_{i=1}[s_{i-1},s_i): N=1,2,\ldots\}$ satisfies that for any given $\rho>0$, $\int_T {\rm P}(|\mu_d(t) - \mu^N_d(t)|\geq \rho)d\mathcal{L}(t) \rightarrow 0$ as $N \rightarrow \infty$
\end{itemize}

Conditions C1 and C2 are similar to conditions 1-2 in Sun et al. (2015). Condition C1 states that $\{\mu_d(t): t\in T\}$ is a smooth process that does not degenerate at both points $-\Delta$ and $\Delta$. It is naturally holds when $\{\mu_d(t): t\in T\}$ is a continuous random process, which ensures that the inequality between $z(t)$ and $z^N(t)$ only occurs with a small chance when $|\mu^N_d(t) - \mu_d(t)|$ is small, where $z^N(t)=\sum^N_{i=1}z(t^*_i)I(s_{i-1}\leq t < s_i)$. Condition C2 requires that the partition $T=\cup^N_{i=1}[s_{i-1},s_i)$ should produce roughly homogeneous subintervals so that the decision at the center point $t^*_i$ can be a good representation of the decision process on subinterval $[s_{i-1}, s_i)$. Then, we will need a lemma of Sun et al. (2015) (Lemma 2). We re-state the result.

{\lemm \label{lem1} Under conditions C1 and C2, $\lim_{N \rightarrow \infty} \int_T {\rm P}(z(t) \neq z^N(t))d\mathcal{L}(t) = 0$. }

{\bf{Proof of Theorem~\ref{theo2}}}. Let $S_k(t)= {\rm P}(z(t)=k\mid \mathcal{D})$, $k=0,1,2$. According to the definition of FDR$_{\textup{I}}$, the FDR$_{\textup{I}}$ level of procedure \eqref{mFDRtest} is
\begin{eqnarray*}
\textup{FDR}_{\textup{I}} &\leq& \E\left\{\frac{1}{\mathcal{L}(R_1\cup R_2)\vee 1}\int^1_0 S_0(t)I(\delta^{(I)}(t)\neq 0)d\mathcal{L}(t)\right\}\\
&=& \E\left\{\frac{1}{\mathcal{L}(R_1\cup R_2)\vee 1}\sum^N_{i=1}I(\delta^{(I)}(t^*_i)\neq 0)\int^{s_i}_{s_{i-1}} S_0(t)d\mathcal{L}(t)\right\}\\
&=& \E\left\{\frac{1}{N(\mathcal{L}(R_1\cup R_2)\vee 1)}\sum^N_{i=1}I(\delta^{(I)}(t^*_i)\neq 0) S_0(t^*_i)\right\} + A_N,
\end{eqnarray*}
where $A_N = \E[\{\mathcal{L}(R_1\cup R_2)\vee 1\}^{-1}\sum^N_{i=1} I(\delta^{(I)}(t^*_i)\neq 0)\int^{s_i}_{s_{i-1}} (S_0(t^*_i)-S_0(t))d\mathcal{L}(t)]$.

Further, let $S^N_k(t)={\rm P}(z^N(t)=k \mid \mathcal{D})$, $k=0,1,2$. Note that $\E|S^N_k(t)-S_k(t)| = {\rm P}(z^N(t)=k, z(t)\neq k) + {\rm P}(z(t)=k, z^N(t)\neq k)$. Then, an application of Lemma~\ref{lem1} yields that
\begin{eqnarray*}
A_N &=& \E\left\{\frac{1}{\mathcal{L}(R_1\cup R_2)\vee 1} \int^1_0 I(\delta^{(I)}(t)\neq 0)(S^N_0(t)-S_0(t))d\mathcal{L}(t)\right\}\\
&\leq& \int^1_0 \E[I(\delta^{(I)}(t)\neq 0)\{S^N_0(t)-S_0(t)\}]d\mathcal{L}(t)\\
&\leq& 2\int^1_0 {\rm P}(z(t) \neq z^N(t))d\mathcal{L}(t)\rightarrow 0,
\end{eqnarray*}
where the second inequality follows from the fact that $\{\mathcal{L}(R_1\cup R_2)\vee 1\}^{-1} \leq 1$. Since the proposed procedure guarantees that
\begin{equation*}
\frac{1}{N(\mathcal{L}(R_1\cup R_2)\vee 1)}\sum^N_{i=1} I(\delta^{(I)}(t^*_i)\neq 0) S_0(t^*_i) \leq \alpha
\end{equation*}
for all realization of $\mathcal{D}$, the FDR$_{\textup{I}}$ is controlled at level $\alpha$ asymptotically. \hfill $\fbox{}$

\subsection*{Appendix B. Derivation of equations \eqref{meanbar}}

Note that $f_{\ve{\Theta}}(Y,\ve{\tilde{\mu}},\vesub{\tilde{\mu}}{d})=f_{\ve{\Theta}}(Y)f_{\ve{\Theta}}(\ve{\tilde{\mu}}, \vesub{\tilde{\mu}}{d}\mid \mathcal{D})$, where $f_{\ve{\Theta}}(Y)$ does not contain any information about $\ve{\tilde{\mu}}$ and $\vesub{\tilde{\mu}}{d}$. Hence, we have
\begin{eqnarray*}
f_{\ve{\Theta}}(\ve{\tilde{\mu}}, \vesub{\tilde{\mu}}{d}\mid \mathcal{D}) &\propto& f_{\ve{\Theta}}(Y,\ve{\tilde{\mu}}, \vesub{\tilde{\mu}}{d})\\
&\propto&  \phi(\ve{\tilde{\mu}}\mid \mathbf{0},\vesub{K}{n}) \phi(\vesub{\tilde{\mu}}{d}\mid \mathbf{0},\vesub{\Gamma}{n_1})\prod^{n_1}_{i=1}\phi(Y_{1i}\mid \mu(t_{1i})+\mu_d(t_{1i}), \sigma^2) \prod^{n_2}_{i=1}\phi(Y_{2i}\mid \mu(t_{2i}), \sigma^2).
\end{eqnarray*}
Then, it is straightforward to know that
\begin{eqnarray*}
f_{\ve{\Theta}}(\vesub{\tilde{\mu}}{d}\mid \mathcal{D}) &=& \int f_{\ve{\Theta}}(\ve{\tilde{\mu}}, \vesub{\tilde{\mu}}{d}\mid \mathcal{D}) d\ve{\tilde{\mu}}\\
&\propto& \exp\left\{-\frac{1}{2\sigma^2} (\vesub{\tilde{\mu}}{d}-\vesup{A}{-1}\ve{b})^T \ve{A} (\vesub{\tilde{\mu}}{d}-\vesup{A}{-1}\ve{b})\right\},
\end{eqnarray*}
where $\ve{A} = \sigma^2\vess{\Gamma}{n1}{-1}+\vesub{I}{n1}-\vesub{\Sigma}{11}$ and $\ve{b}=(\vesub{I}{n1}-\vesub{\Sigma}{11})\vesub{Y}{1}-\vesub{\Sigma}{12}\vesub{Y}{2}$. It implies that $\vesub{\tilde{\mu}}{d} = \vesup{A}{-1}\ve{b} + \vesub{\epsilon}{1}$ with $\vesub{\epsilon}{1} \sim N(\mathbf{0}, \sigma^2\vesup{A}{-1})$. On the other hand, note that $(\vess{\tilde{\mu}}{d}{T}, \vess{\mu}{d}{NT})^T$ follows a multivariate normal distribution with mean zero and covariance matrix $\ve{\Gamma}$, where
\begin{equation*}
\ve{\Gamma} = \left(
           \begin{array}{cc}
             \vesub{\Gamma}{n1} & \vesup{\Psi}{T}(T^*) \\
             \ve{\Psi}(T^*) & \vesub{\Gamma}{N} \\
           \end{array}
         \right).
\end{equation*}
Thus, we have $\vess{\mu}{d}{N} = \ve{\Psi}(T^*)\vess{\Gamma}{n1}{-1}\vesub{\tilde{\mu}}{d} + \vesub{\epsilon}{2}$ with $\vesub{\epsilon}{2} \sim N(\mathbf{0}, \vesub{\Gamma}{N} - \ve{\Psi}(T^*)\vess{\Gamma}{n1}{-1}\vesub{\Psi}{T}(T^*))$. Consequently, $\vess{\mu}{d}{N} = \ve{\Psi}(T^*)\vess{\Gamma}{n1}{-1}\vesup{A}{-1}\ve{b} + \ve{\Psi}(T^*)\vess{\Gamma}{n1}{-1}\vesub{\epsilon}{1} + \vesub{\epsilon}{2}$, so the conditional distribution of $\vess{\mu}{d}{N}$ given $\mathcal{D}$ is a multivariate normal distribution with mean $\ve{\Psi}(T^*)\vess{\Gamma}{n1}{-1}\vesup{A}{-1}\ve{b}$ and covariance matrix $\sigma^2\ve{\Psi}(T^*)\vess{\Gamma}{n1}{-1}\vesup{A}{-1}\vess{\Gamma}{n1}{-1}\vesup{\Psi}{T}(T^*) + \vesub{\Gamma}{N} - \ve{\Psi}(T^*)\vess{\Gamma}{n1}{-1}\vesup{\Psi}{T}(T^*)$, i.e., $\vess{\mu}{d}{N}\mid \mathcal{D}\sim N(\ve{\bar{\mu}}, \ve{\Lambda})$.

\end{document}